\documentclass[letterpaper,twoside,12pt]{article}

\usepackage{url}
\usepackage{comment}
\usepackage{footmisc}

\usepackage{tocloft}
\usepackage{rotating}
\usepackage{amsmath}
\usepackage[
top    = 2.75cm,
bottom = 2.50cm,
left   = 2.65cm,
right  = 2.65cm]{geometry}
\usepackage{lscape}
\usepackage{pdflscape}

\usepackage{multirow}
\usepackage{booktabs}
\usepackage{subcaption}
\usepackage{amsmath}
\usepackage{amsfonts}
\usepackage{MnSymbol}
\usepackage{wasysym}
\usepackage{dcolumn,booktabs}
\newcolumntype{d}[1]{D{.}{.}{#1}}

\usepackage{hyperref}
\usepackage{natbib}
\usepackage{hypernat}
\hypersetup{colorlinks=true,
            citecolor=[rgb]{0,0.2,0.8},
            linkcolor=[rgb]{0,0.2,0.8},
            urlcolor =[rgb]{0,0,0}}
\usepackage{fancyhdr}                   
\usepackage{fancyvrb}
\usepackage{lipsum}
\usepackage{caption}
\usepackage{setspace}
\usepackage{relsize}
\usepackage{graphicx}
\usepackage{xfrac}
\usepackage{floatrow}
\usepackage{titlesec}

\titlespacing\section{0pt}{6pt plus 0pt minus 0pt}{6pt plus 0pt minus 0pt}
\titlespacing\subsection{0pt}{6pt plus 2pt minus 2pt}{0pt plus 2pt minus 2pt}

\newif\iflogvar
\logvartrue

\newif\ifblankver
\blankverfalse

\DeclareMathOperator*{\argmax}{argmax} 
\DeclareMathOperator*{\argmin}{argmin} 

\hypersetup{colorlinks=true,
            citecolor=[rgb]{0,0.2,0.8},
            linkcolor=[rgb]{0,0.2,0.8},
            urlcolor =[rgb]{0,0,0}}

\usepackage{color}
\definecolor{blue}{rgb}{0,0.2,0.8}

\ifblankver
\else
\fi
\ifblankver

\title{Can Volatility Solve the Naive Portfolio Puzzle?}
\vspace{0.5cm}
\author{}

\else

\title{Can Volatility Solve the Naive Portfolio Puzzle?\footnote{\scriptsize 
We thank Caitlin Dannhauser, Jes\'{u}s Fern\'{a}ndez-Villaverde, Alejandro Lopez-Lira, Rabih Moussawi, Michael Pagano, Nikolai Roussanov, Paul Scanlon, Frank Schorfheide, John Sedunov, Raman Uppal, and Raisa Velthuis for helpful comments. Christopher Antonello provided diligent research assistance. 
There are no conflicts of interest to disclose. 
A technical appendix can be found at \url{http://michael-curran.com/research/volatility_appendix.pdf}.
}
}\vspace{0.5cm}

\author{\href{http://www.michael-curran.com}{Michael Curran}\footnote{\scriptsize 
Corresponding author. 
Email: michael.curran@villanova.edu; Phone: (+1) 610-519-8867
\newline Address: Economics Dept., Villanova School of Business, Villanova University, 800 E Lancaster Ave, PA 19085, USA.} \\
Villanova University 
\and 
Patrick O'Sullivan\footnote{\scriptsize
Email: patrick.osullivan@schroders.com 
\newline Address: Schroders, 1 London Wall Place, London, UK.} \\
Schroders Investment Management
\and \href{https://sites.google.com/site/ryanzalla/}{Ryan Zalla}\footnote{\scriptsize Email: rzalla@sas.upenn.edu; Phone: (+1) 412-759-5032 \newline Address: Economics Dept., University of Pennsylvania, 133 South 36th Street, Philadelphia, PA 19104, USA.} \\
 University of Pennsylvania \vspace{.4cm}} 

\fi

\date{\small \today \\ \vspace{-0.5cm}} 

\begin{document}

\begin{titlepage}

\thispagestyle{empty} 

\begin{onehalfspace}
\maketitle
\end{onehalfspace}
\vspace{-0.5cm}
\begin{doublespace}
\begin{abstract}
We investigate whether sophisticated volatility estimation improves the out-of-sample performance of mean-variance portfolio strategies relative to the naive 1/N strategy. The portfolio strategies rely solely upon second moments. Using a diverse group of portfolios and econometric models across multiple datasets, most models achieve higher Sharpe ratios and lower portfolio volatility that are statistically and economically significant relative to the naive rule, even after controlling for turnover costs. Our results suggest benefits to employing more sophisticated econometric models than the sample covariance matrix, and that mean-variance strategies often outperform the naive portfolio across multiple datasets and assessment criteria.
\end{abstract}
\end{doublespace}

%
\indent\small \textbf{Keywords:} mean-variance, naive portfolio, volatility \\

\indent\small \textbf{JEL:} G11, G17
\end{titlepage}

\doublespacing

\newpage \setcounter{page}{1}

\setlength{\baselineskip}{1.1\baselineskip} 
\setlength{\parskip}{0pt plus0pt}

\section{Introduction}

\noindent{}Since~\citet{markowitz1952} introduced mean-variance strategies, their out-of-sample performance has been criticized. One reason for weak performance is estimation error in the mean return (\citealp{merton1980estimating}; \citealp{chopra2013effect}). Variance strategies, using only second moments, avoid the pitfall of expected returns estimation. \citet{demiguel2009optimal} found the difference in performance of mean-variance strategies relative to an equally diversified portfolio to be statistically insignificant. We call this the \textit{naive} diversification puzzle. While previous work such as~\citet{demiguel2009optimal} employs robust estimation procedures to reduce parameter errors, they use sample covariance estimates.
Few papers in the naive portfolio literature have pursued improved estimation of the volatility of returns.\footnote{Instead of the portfolio strategy, our innovation explores a wide variety of \textit{econometric} models. \citet{demiguel2009optimal} find that the minimum-variance portfolio, though performing well relative to other portfolio strategies, significantly beats the 1/N strategy for only 1 in 7 of their datasets. \citet{jagannathan2003} and~\citet{kirby2012s} innovate on the portfolio strategy, illustrating that short-sale constrained minimum-variance strategies and volatility timing strategies enhance performance.}\textsuperscript{,}\footnote{We consider a wide range of mostly parametric econometric models. Non-parametric models using higher-frequency data~\citep{demiguel2013improving} and shrinkage approaches~\citep{ledoit2017nonlinear} also improve the accuracy of estimation. Daily frequency option-implied volatility reduces portfolio volatility, but never statistically significantly improves the Sharpe ratio relative to the 1/N strategy~\citep{demiguel2013improving}. Although \citet{johannes2014sequential} account for both estimation risk and time-varying volatility through eight variations of a similar class of constant and stochastic volatility models, we expand to more varied classes of volatility types with 14 econometric models. Initial investigations reveal our results to be at least as strong as~\citet{ledoit2017nonlinear}. 
} Our contribution is that a variety of econometric models of volatility improve performance relative to the naive portfolio strategy. That is, relative to sample covariance estimation, our study suggests considerable performance gains from employing modern econometric models for estimating volatility in portfolio construction.

We investigate whether sophisticated volatility estimation improves the out-of-sample performance of mean-variance portfolio strategies depending only on conditional variance-covariance matrices (variance strategies) relative to the naive 1/N strategy. Using a diverse set of fourteen econometric models, we 
apply three portfolio strategies -- minimum-variance, constrained minimum-variance, and volatility-timing -- to six empirical datasets of weekly and monthly returns; we also include the tangency portfolio. 
We assess performance using three criteria: (i) the Sharpe ratio, (ii) the Sharpe ratio adjusted for turnover (trading) costs, and (iii) the standard deviation of returns (portfolio volatility). Overall, we show that variance strategies perform consistently well out-of-sample.

Our first contribution is that all four portfolio strategies, regardless of the econometric model used for estimating covariance, achieve superior out-of-sample performances relative to the naive benchmark. This assertion challenges the literature that has rarely reported superior minimum-variance performances relative to the equally-weighted portfolio \citep{demiguel2009optimal}.\footnote{Our econometric estimation strategies yield improvements beyond the period and frequency differences.} Specifically, we find that all four portfolio strategies estimated using all fourteen econometric models perform \textit{at least as well} as the naive benchmark, and only rarely underperform it. These underperformances are mainly concentrated in the Fama-French 3-factor dataset. If we discard this dataset, 
then all four portfolio strategies estimated using twelve of the fourteen econometric models would \textit{weakly dominate} the naive benchmark.\footnote{A portfolio strategy, whose covariance is estimated using a given econometric model, \textit{weakly dominates} the naive benchmark if, for each performance criterion, the portfolio strategy performs at least as well as the naive benchmark across all datasets and performs significantly better in at least one dataset.}


In general, the minimum-variance strategy, with or without short-sale constraints, achieves higher Sharpe ratios, lower turnover costs, and lower portfolio volatility across the majority of datasets. Wherever these strategies do not outperform the naive rule, we often fail to reject the null hypothesis that their performances are identical to the naive rule. 
Likewise, we find evidence that volatility-timing strategies achieve higher Sharpe ratios and lower turnover costs but exhibit comparable portfolio volatility to the naive rule. The tangency portfolio does reasonably well in certain datasets. Relative to the other strategies, however, its performance is lackluster, which we attribute to estimation error in expected returns.


Our second contribution is to identify pairings of volatility models and portfolio strategies that perform consistently and significantly well across datasets relative to the naive benchmark. Multivariate GARCH models, particularly the constant conditional correlation (CCC), \textit{weakly dominate} the naive rule when applied to minimum-variance and constrained minimum-variance strategies. Similarly, the realized covariance (RCOV) model weakly dominates the naive rule when applied to the volatility-timing strategy. In the tangency portfolio, although the RCOV model achieves higher Sharpe ratios and lower portfolio volatility relative to the naive strategy than other econometric models, it exhibits higher portfolio volatility relative to the naive rule in data on international equities. Nonetheless, many other econometric models, such as the multivariate GARCH models, weakly dominate the naive strategy. Even econometric models such as the regime-switching vector autoregression (RSVAR) and exponentially-weighted moving-average (EWMA), which perform worst in each of the four porfolio models, still perform at least as well as the naive benchmark across every dataset except the Fama-French 3-factor.


Our third contribution is to compare the performance of econometric models relative to the naive benchmark using each assessment criterion in \textit{isolation}. That is, across portfolio strategies and datasets, which econometric models consistently achieve the highest Sharpe ratios? Which econometric models achieve the lowest turnover costs? Which econometric models achieve the lowest portfolio volatilities? First, the combined parameter (CP) and realized covariance (RCOV) models achieve significantly higher Sharpe ratios than the naive benchmark. Second, the RCOV and a variety of GARCH models produce significantly higher Sharpe ratios than the naive rule after adjustment for turnover costs. Finally, the exponentially-weighted moving-average (EWMA), multivariate stochastic volatility (MSV), and RCOV models exhibit significantly lower portfolio volatility than the naive rule. These performances are economically and statistically significant: relative to the naive strategy, we achieve 30\% higher Sharpe ratios and 9\% lower portfolio volatility.\footnote{For each portfolio strategy, we average the Sharpe ratios and portfolio volatility resulting from all fourteen econometric models across all six datasets. Then we average Sharpe ratio and portfolio volatility across all four portfolio strategies.} 



Our paper exploits recent computational developments to estimate the covariance matrix 
using multivariate, nonlinear, non-Gaussian econometric models.\footnote{Our study benefits from incorporating recent advances in the computation of several models as in~\citet{vogiatzoglou2017dynamic}, \citet{chan2018bayesian}, and~\citet{kastner2019sparse}. To reduce run-time, we employ 
fast, low-level languages, e.g., C++, that we program in parallel with hyperthreading and execute on 
clusters.} 
We extend the set of models in~\citet{wang2015hedging} beyond 
GARCH and discrete regime-switching models to smooth multivariate stochastic volatility 
and non-parametric realized volatility models. 

We draw two conclusions from our results. First, relative to sample covariance estimation, our study suggests considerable performance gains from employing modern econometric models for estimating volatility in portfolio construction.  
Second, 
variance strategies consistently perform well relative to the naive strategy. 



Our study builds on the literature of naive diversification. 
Mean-variance models struggle to compete with naive diversification in out-of-sample return performance. Considering multiple mean-variance strategies including ones accounting for parameter errors, 
\citet{demiguel2009optimal} 
find that no portfolio strategy 
consistently outperforms naive diversification. 
One cause for weak performance is that while mean-variance strategies based on the minimum-variance portfolio outperforms naive diversification, portfolio turnover costs negates these benefits~\citep{kirby2012s}. To address the turnover issue, the authors  
propose two new approaches that reduce portfolio turnover, finding that the mean-variance strategies outperform the naive strategy out-of-sample. We employ their volatility-timing strategy.  
Combining naive diversification with other conventional portfolio 
strategies can also enhance portfolio performance~\citep{tu2011markowitz}. We investigate combining parameters from the variance-covariance matrix estimates with portfolio strategy weights.

Studies have increased the sophistication of parameter estimation since \citet{demiguel2009optimal}, which was based on rolling window sample estimates. Performances of mostly GARCH-based estimators were lackluster~\citep{trucios2019covariance}.\footnote{Using a shorter time-sample across one dataset with a larger portfolio, they do not consider vector autoregression, vector error correction for non-stationarity, or either regime-switching or stochastic volatility models, which are computationally challenging and account for observed nuances of time-varying volatility.} Non-linear shrinkage estimators~\citep{ledoit2017nonlinear} and estimation strategies for large portfolios~\citep{ao2019approaching} can improve performance.\footnote{Preliminary evidence suggests that our results are at least as strong relative to the naive portfolio as what~\citet{ledoit2017nonlinear} find. Direct comparisons are more complicated in~\citet{ao2019approaching}. Relative to the naive portfolio, initial experiments suggest that their MAXSER estimator performs better than our econometric models do in some comparisons, but that our models do better in most empirical comparisons.} Implied volatility and skewness~\citep{demiguel2013improving}, 
vector autoregression~\citep{demiguel2014stock},  
and portfolio constraints (\citealp{demiguel2009generalized}; \citealp{kourtis2012parameter}; \citealp{behr2013portfolio}) 
also enhances portfolio performance.\footnote{To isolate one study by~\citet{demiguel2009generalized}, our paper employs improved econometric methods rather than more sophisticated portfolio constraints. Although not directly comparable, preliminary investigations reveal that our models improve performance relative to the naive portfolio by a greater ratio than~\citet{demiguel2009generalized} in terms of Sharpe ratios, portfolio volatility, and turnover costs.}  
%
Recent papers find comparably poor performance using sophisticated hedging strategies to beat naive one-to-one hedging~\citep{wang2015hedging}, provide behavioral evidence of naive choice strategies (\citealp{benartzi2001}; \citealp{huberman2006offering};  
 \citealp{giorgi2018naive}; \citealp{gathergood2019naive}), and emphasize the importance of volatility~\citep{moreira2017volatility,moreira2019volatility}. 
Consistent with 
the latter,
our paper finds that volatility-timing strategies generate higher Sharpe ratios yet exhibit equal volatility to the naive portfolio. We explore improved volatility estimation as a source of out-of-sample performance. 
Rather than consider an investor allocating wealth across individual assets, our paper differs from many papers by focusing on the problem of allocating wealth across portfolios. 
We apply our methods to six different empirical datasets with $N\leq11$ portfolio choices. Many retail investors and some institutional investors trade a small number of index portfolios rather than a large number of individual stocks. Evidence using over half a million individuals in over six hundred 401(k) plans indicates that participants tend to use three or four funds and allocate their contributions equally across the funds~\citep{huberman2006offering}. 
Fortunately, our multiple datasets are broad and varied. For instance, we cover datasets allowing investors to diversify across international equities in the form of national stock market aggregate indices, across entire sector and industry indices, across expansive Fama-French portfolios, and across other encompassing indices based on size/book-to-market and momentum portfolios. The best application for our methods from a practical standpoint, therefore, is to an investor holding multiple mutual funds in equities. 

Our benchmark is the \textit{naive} diversification rule,  allocating a fraction $1/N$ of wealth to each of $N$ choices available for investment at each rebalancing date. 
Three reasons justify the naive rule as a benchmark. First, implementation is easy as it relies on neither optimization nor estimation of the moments of returns. Second, investors use such simple rules for allocating their wealth across investments (\citealp{benartzi2001}; \citealp{huberman2006offering}; 
\citealp{baltussen2011}; \citealp{giorgi2018naive}; \citealp{gathergood2019naive}). Third, the naive rule consistently outperforms mean-variance strategies 
(\citealp{demiguel2009optimal}; \citealp{duchin2009}; \citealp{pflug2012}). Without needing to estimate the 1/N weights, the variance of parameter estimation is zero, and thus the mean square error of the naive portfolio weights is simply the square of the bias.\footnote{Letting $\hat{\mathbf{w}}$ denote our estimate of the optimal vector of portfolio weights $\mathbf{w}$, the MSE bias-variance decomposition from econometrics is $MSE(\hat{\mathbf{w}})=Var(\hat{\mathbf{w}})+Bias^2(\hat{\mathbf{w}},\mathbf{w})$, where $Bias(\hat{\mathbf{w}},\mathbf{w})=\hat{\mathbf{w}}-\mathbf{w}$.} The naive strategy therefore proxies as a challenging rival for mean-variance strategies to outperform.

\section{Econometric Models}\label{sec:metrics}

We benchmark our model choices to~\citet{wang2015hedging}, who consider out-of-sample performance of hedging strategies.\footnote{While we attempt to cover the broad classes of econometric models, our set of econometric models is not exhaustive. For instance, we omit the shrinkage estimators of~\citet{hafner2012estimation} and~\citet{ledoit2003improved,ledoit2017nonlinear}. Although these and other econometric models are interesting, our study is the most expansive in its coverage of econometric models.} We extend their analysis from bivariate to multivariate random variables where the choice set is $N>2$. For each of their econometric models, they experiment with multiple estimators. Initial investigations reveal that heterogeneity in econometric models matters more than heterogeneity in estimators for yielding differing means, covariance estimates, and portfolio weights. We therefore expand the set of econometric models, controlling for different real world features in the data.\footnote{Implementation details are relegated to the online appendix.} 
Letting $M$ be the estimation window length with $T$ out-of-sample investment periods, we use a rolling window approach with $T+M$ returns for $N$ choices. 
Picking ten-year rolling windows, in line with~\citet{demiguel2009optimal}, $M$ will be set by the frequency of the data, e.g., $M=520$ for weekly data and $M=120$ for monthly data. We choose $T=1$, which corresponds to one week or month ahead, reduces compute time, and allows us to compare our results with the literature, e.g.,~\citet{demiguel2009optimal}. With $P$ total periods in our data, let $\{\hat{\Sigma}_t^j{}\}_{t=M}^{P-T}$ denote the conditional estimate of the variance-covariance matrix of returns for investment period $t$ based on econometric model $j$. Similarly, we define the conditional estimate of the expected return over period $t$ given model $j$ by $\{\hat{\boldsymbol{\mu}}_t^j\}_{t=M}^{P-T}$. 
Mean-variance strategies are defined by the first two conditional moments of the return for period $t$ so that $\{(\hat{\boldsymbol{\mu}}_t^j,\hat{\Sigma}_t^j)\}_{t=M}^{P-T}$ defines the sequence of mean-variance strategies over the $T$ out-of-sample investment periods with respect to model $j$. Section~\ref{sec:pfm} details the mean-variance strategies. 

\subsection{Sample Covariance}\label{subsec:cov}

\noindent{}
Sample-based in-sample estimation of the variance-covariance matrix is standard in the literature on the naive diversification puzzle (\citealp{demiguel2009optimal}; \citealp{fletcher2011optimal}; \citealp{tu2011markowitz}; \citealp{kirby2012s}). Some studies examine improved estimation but typically use a small set of models~\citep{demiguel2013improving}. The sample covariance matrix (Cov) equally weights past observations.

\subsection{Exponentially Weighted Moving Average}\label{subsec:ewma}

\noindent{}The recent past might be more informative for estimating the variance-covariance matrix, motivating our first refinement, the exponentially weighted moving average (EWMA) model. The EWMA model suggested by RiskMetrics places decaying weight on the past. 

\subsection{Vector Autoregression}\label{subsec:var}

\noindent{}To exploit dependence along the cross-section and time-dimension (serial), we estimate a vector autoregression (VAR). 
 We typically find two lags to be optimal at weekly and monthly frequencies, i.e., we estimate a VAR(2). 

\subsection{Vector Error Correction}\label{subsec:vec}

\noindent{}To account for potential cointegration between the variables, we estimate a parsimonious vector error correction (VEC) model. 
We compute the number of cointegrating relations in the system following the Johansen trace test~\citep{johansen1988statistical,johansen1991estimation} and employ the variance-covariance matrix estimated from the VEC model. 

\subsection{BEKK-GARCH and Asymmetric BEKK-GARCH}\label{subsec:bekk} 

\noindent{}The volatility of financial return data varies over time. General Autoregressive Conditional Heteroscedasticity (GARCH) 
models the evolution of volatility as a deterministic function of past volatility and innovations. 
%
 Our first multivariate GARCH specification is BEKK-GARCH, modeling causalities of variances by allowing the conditional variances of one variable to depend on lagged values of another. Empirically, BEKK is general but easy to estimate. Relaxing symmetry, we also allow positive and negative shocks of equal magnitude to have different effects on conditional volatility by employing Asymmetric BEKK (ABEKK). 
We allow 
one symmetric innovation when estimating BEKK, and one symmetric innovation and one asymmetric innovation when estimating ABEKK. 

\subsection{Conditional Correlation: Constant, Dynamic, \& Asymmetric}\label{subsec:ccc}

\noindent{}BEKK and ABEKK suffer from the curse of dimensionality which renders them computationally infeasible for investors allocating capital across a large set of investment choices. We therefore also estimate a constant conditional correlation (CCC) model. The CCC model is a multivariate GARCH model, where all conditional correlations are constant and conditional variances are modeled by univariate GARCH processes.

The CCC model benefits from almost unrestricted applicability for large systems of time series, but fails to account for increases in correlation during financial crises. 
Dynamic conditional correlation (DCC) permits time-varying correlation
. 
Without accounting for dynamics of asymmetric effects (ADCC), however, DCC cannot distinguish between the effect of past positive and negative shocks on the future conditional volatility and levels. 
 We allow 
one symmetric innovation when estimating CCC and DCC, and one symmetric innovation and one asymmetric innovation when estimating ADCC. 

\subsection{Copula-GARCH}\label{subsec:copula}

\noindent{}The assumption of multivariate normality is often questioned in practical applications. For instance, shocks that affect Apple may also affect Microsoft. Each company may experience similar nonlinear extreme events, hence exhibiting tail dependence. A portfolio manager who assumes multivariate normality will underestimate the frequency and magnitude of rare events. Such underestimation may hurt the portfolio's performance.

Modeling multivariate dependence among stock returns without assuming multivariate normality has become popular in the 21st century. Copulas are functions that may be used to bind univariate marginal distributions to produce a multivariate distribution. 
Parameters can vary over time as an autoregression in a copula-GARCH model. Copulas have become the standard tools for modeling multivariate dependence among stock returns without assuming multivariate normality with many general applications in finance. 

\subsection{Regime-Switching Vector Autoregression}\label{subsec:rsvar}

\noindent{}To account for bull and bear phases of the market, we estimate a discrete time-varying parameter model in the form of a regime-switching VAR (RSVAR) as in~\citet{chan2018bayesian}. 
We choose two regimes 
and we set our lag length at 2 for parsimony. 

\subsection{Multivariate Stochastic Volatility}\label{subsec:msv}

\noindent{}Another nonlinear state-space model that allows for heteroscedasticity is the computationally challenging multivariate stochastic volatility model (MSV). Unlike GARCH models, volatility is stochastic. We adapt our method from~\citet{kastner2019factorstochvol,kastner2019sparse}. 

\subsection{Realized Volatility}\label{subsec:rcov}

\noindent{}Our final econometric model is a non-parametric model: realized volatility (RCOV). To augment our group of econometric models from consisting of purely parametric models, we choose this model, although a backwards looking one, because of its popularity in practice and in the financial volatility literature. 

\section{Portfolio Strategies}\label{sec:pfm}

\noindent{}
To avoid the significant issues with estimating mean returns \citep{merton1980estimating} and the large impact of errors in the mean vector on out-of-sample performance \citep{chopra2013effect}, we concentrate on portfolio strategies that depend only on the covariance matrix. 

\subsection{Naive Diversification}

\noindent{}With $N$ investment choices, the portfolio held over investment period $t$, $\mathbf{w}_t^{NV}$, is given by
\begin{equation}
\mathbf{w}_t^{NV} = (1/N,\ldots,1/N)	\quad	\forall t.
\label{eq:nv}
\end{equation}

\subsection{Minimum-Variance Portfolio}

\noindent{}The minimum-variance portfolio (MVP) for investment period $t$, $\mathbf{w}_t^{MVP}$, minimizes conditional portfolio variance. For each econometric model $j$ discussed in Section~\ref{sec:metrics}, we calculate the minimum-variance portfolio by
\begin{equation}
\mathbf{w}_{t,j}^{MVP} = \argmin_{\mathbf{w}\in\mathbb{R}^N|\mathbf{w}'\mathbf{1}=\mathbf{1}}{\mathbf{w}\hat{\Sigma_t^j}{}\mathbf{w'}}.
\label{eq:mvp}
\end{equation}
For each econometric model, the conditional estimate of the covariance matrix over period $t$ is used as an input to find the minimum-variance portfolio.

\subsection{Constrained Minimum-Variance Portfolio}

\noindent{}The constrained minimum-variance portfolio (con-MVP) for investment period $t$, $\mathbf{w}_t^{Con-MVP}$, minimizes conditional portfolio variance subject to no short selling; con-MVP improves performances (\citealp{jagannathan2003}; \citealp{demiguel2009optimal}). For each econometric model $j$ discussed in Section~\ref{sec:metrics}, we calculate the minimum-variance portfolio by
\begin{equation}
\mathbf{w}_{t,j}^{Con-MVP} = \argmin_{\mathbf{w}\in\mathbb{R}^N|\mathbf{w}'\mathbf{1}=\mathbf{1},\mathbf{w}\geq0}{\mathbf{w}\hat{\Sigma_t^j}{}\mathbf{w'}}.
\end{equation}
For each econometric model, we use the conditional estimate of the covariance matrix over period $t$ as an input to find the constrained minimum-variance portfolio.

\subsection{Volatility-Timing Strategies}

\noindent{}The volatility-timing strategy 
ignores off-diagonal elements of the covariance matrix, i.e., assumes all pair-wise correlations are zero~\citep{kirby2012s}. The minimum-variance portfolio given covariance matrix $\Sigma$ is  $w_i^{VT}=\frac{1/\Sigma_{ii}}{\sum_{i=1}^N{1/\Sigma_{ii}}}$. We similarly define the volatility-timing (VT) strategy given conditional estimate of the covariance matrix $\hat{\Sigma}_t^j$ by
\begin{equation}
\left(w_{t,j}^{VT}\right)_i = \frac{1/\left(\hat{\Sigma}_t^j\right)_{i,i}}{\sum_{i=1}^N{1/\left(\hat{\Sigma}_t^j\right)_{i,i}}} \qquad i=1,\ldots,N.
\end{equation}

\subsection{Tangency Portfolio}

\noindent{}While our main focus is on minimum-variance portfolio strategies, we also include the tangency portfolio (TP) for illustrative purposes. The TP with respect to econometric model $j$, $\mathbf{w}_{t,j}^{TP}$, is given by
\begin{equation}
\mathbf{w}_{t,j}^{TP} = \argmax_{\mathbf{w}\in\mathbb{R}^N|\mathbf{w}'\mathbf{1}=\mathbf{1}}{\frac{\mathbf{w}\hat{\boldsymbol{\mu}}_t^j}{\mathbf{w}\hat{\Sigma}_t^j\mathbf{w}'}}.
\label{eq:tp}
\end{equation}

\subsection{Combined Parameter Strategy}\label{subsec:cp}

\noindent{}We combine portfolios by inputting the arithmetic average over the econometric estimates of the covariance matrix into the portfolio optimization strategies. 
 We form a combined parameter estimate of the covariance matrix $\hat{\Sigma}$ by equally weighting estimates of the covariance matrix from each of the thirteen econometric models. We use the combined parameter estimate for $\hat{\Sigma}$ in each of~\eqref{eq:mvp}--\eqref{eq:tp} to get four combined parameter (CP) portfolios. Taking the example of the minimum-variance portfolio, using $\hat{\Sigma}^{comv}$ as the arithmetic average of the covariance matrices across the thirteen econometric models, our CP strategy is
\begin{equation}
\mathbf{w}_{t,j}^{MVP,comv} = \argmin_{\mathbf{w}\in\mathbb{R}^N|\mathbf{w}'\mathbf{1}=\mathbf{1}}{\mathbf{w}\hat{\Sigma}_t^{comv}\mathbf{w'}}.
\label{eq:comv}
\end{equation}
 %
%
 Our CP strategy is motivated by the finding that combining different hedging forecasts leads to more consistent hedging performance across datasets~\citep{wang2015hedging}. The result echoes the forecasting literature finding that combined models tend to perform more consistently over time than individual models~\citep{stock2003forecasting,stock2004combination}.
 
 In preliminary investigations, we explored two other approaches: 
(i) naively weighting across the thirteen vectors of weights suggested by the portfolio strategy using each econometric model's variance-covariance matrix estimate as an input; 
and (ii) naively weighting across the four weights suggested by the four financial portfolio strategies for a given econometric model.\footnote{First, a variation of our benchmark CP strategy~\eqref{eq:comv},  
%
for each of the strategies~\eqref{eq:mvp}--\eqref{eq:tp}, we examine the corresponding portfolio given by naive investments across the thirteen portfolios with respect to each of the econometric models. More precisely, consider the minimum-variance portfolio. We form a fourteenth portfolio strategy, $w_t^{MVP,com}$, which is equally invested across the thirteen estimates of the true minimum-variance portfolio, i.e.,
\begin{equation*}
\mathbf{w}_t^{MVP,com} = \frac{1}{13}\sum_{j=1}^{13}{\mathbf{w}_{t,j}^{MVP}}.
\end{equation*}
 Second, with respect to each of the econometric models, we examine the corresponding portfolio given by naive investments across the four strategies~\eqref{eq:mvp}--\eqref{eq:tp}. More precisely, consider the VAR econometric model. We form a fifth portfolio strategy, $w_t^{VAR,comp}$ that is equally invested across the four vectors of portfolio weights suggested by inputting the volatility estimates from the VAR model into strategies~\eqref{eq:mvp}--\eqref{eq:tp}, i.e., 
\begin{equation*}
\mathbf{w}_t^{VAR,comp} = \frac{1}{4}\sum_{k=1}^{4}{\mathbf{w}_{t,VAR}^{k}}.
\end{equation*}
\label{fn:cp}} 
 These alternative combined parameter strategies are less relevant for our study. With the first variation, averaging over thirteen weights suggested by the portfolio strategy is less direct than averaging over the variance-covariance matrix estimates. With the second variation, averaging over four portfolio strategies for a given econometric model is similar to that of~\citet{wang2015hedging}. Consider instead the realistic situation that we are unsure of the data generating process underlying the return series. Rather than choosing one econometric model, we benefit from using all the information by hedging equally across the various nuances captured by each of the thirteen econometric models. Results are broadly similar across the three versions of combined parameter strategies. Thus, we report the results from our combined parameter strategy~\eqref{eq:comv}, which we denote CP.

\section{Data}\label{sec:data}

\noindent{}We employ six datasets at weekly and monthly frequencies. Lower frequencies 
smooth out too much volatility and are inappropriate for our study. Higher frequency data generate more accurate estimates of the covariance matrix, but daily and higher frequency data are also troubled with problems such as day-of-the-week effect and asynchronous trading. 
We use value-weighted returns and assess robustness 
to equally-weighted returns. 
We use end-of-period data where possible. When weekly or monthly frequency is unavailable, we scale data geometrically. For instance, to scale returns from daily to weekly frequency, we use $\Pi_{j=1}^{ND}{(1+r_j)^{1/ND}-1}$ where $r_j$ is the daily return and $ND$ denotes the number of trading days in the week. We adopt similar procedures to scale to monthly frequency. 
For the realized covariance (RCOV) model, we use daily data 
to calculate RCOV over each week (month) for weekly (monthly) frequency analysis. We omit data prior to July 1963.\footnote{Standard and Poor's established Compustat in 1962 to serve the needs of financial analysts and back-filed information only for the firms that were deemed to be of the greatest interest to the analysts. The result is significantly sparser coverage prior to 1963 for a selected sample of well performing firms.}

Our choice of datasets is motivated by comparison with previous literature. Our first four datasets closely correspond to datasets 4, 2, 1, and 3 of~\citet{demiguel2009optimal}. Popularly employed in the empirical finance literature, our final two datasets come from the same source as those of dataset 4 by the same authors. 
We choose datasets with a modest number of portfolio choices ($N\approx10$). Choosing to allocate wealth across a small number of index portfolios is in line with evidence on the behavior of many retail investors and some institutional investors~\citep{huberman2006offering}. 
 Rather than examining simulated datasets, such as randomized selections of stocks, we restrict our attention to empirical datasets because our focus is on the econometric model as the source of improvement in performance. 

\subsection{Dataset 1: Fama-French Portfolios}

\noindent{}Our first dataset consists of returns obtained from Wharton Research Data Services.\footnote{Kenneth French provides full description at \url{https://mba.tuck.dartmouth.edu/pages/faculty/ken.french/}} We focus on the three-factor Fama-French portfolio: Small Minus Big, High Minus Low, and Market portfolios. Small Minus Big (SMB) is the average return on three small portfolios minus the average return on three big portfolios. High Minus Low (HML) is the average return on two value portfolios minus the average return on two growth portfolios. 
The Market (MKT) return is the weighted return on all NYSE, AMEX, and NASDAQ stocks from CRSP and is obtained by adding the risk-free return to the excess market return.\footnote{The risk-free (RF) asset is the one-month Treasury bill rate from Ibbotson Associates and proxies the return from investing in the money market. We exclude the risk-free rate from the investor's choice set; therefore, we exclude returns in excess of the risk-free rate.} The benchmark analysis uses value-weighted returns to generate MKT. 
We focus on weekly and monthly frequencies, where we use end-of-period returns for daily and monthly frequencies, and we scale daily data to weekly data as described above. We limit the data to span all US trading days from July 1st, 1963 to December 31st, 2018.

\subsection{Dataset 2: Industry Portfolios}\label{subsec:ind}

\noindent{}We take returns from Kenneth French's website covering ten industries: Consumer-\linebreak{}Discretionary, Consumer-Staples, Manufacturing, Energy, High-Tech, Telecommunications, Wholesale and Retail, Health, Utilities, and Others. The benchmark analysis uses value-weighted returns.\footnote{\label{equal1}We also employ equal-weighting in robustness checks.} We focus on 
weekly and monthly frequencies, where we use end-of-period returns for daily and monthly frequencies, and we scale daily data to weekly data as described above. We limit the data to span all US trading days from July 1st, 1963 to December 31st, 2018.

\subsection{Dataset 3: Sector Portfolios}

\noindent{}This dataset includes returns for eleven value-weighted industry portfolios formed by using the Global Industry Classification Standard (GICS) developed by Standard \& Poor's (S\&P) and Morgan Stanley Capital International (MSCI). We obtained the returns from Bloomberg. The ten industries are Energy, Materials, Industrials, Consumer-Discretionary, Consumer-Staples, Healthcare, Financials, Information-Technology, Telecommunications, Real Estate, and Utilities. The expected returns are based on equity investments. Data are end-of-period returns for weekly and monthly frequencies and span all US trading days from January 2nd, 1995 to December 31st, 2018.

\subsection{Dataset 4: International equity indices}

\noindent{}This dataset includes returns on eight MSCI countries, Canada, France, Germany, Italy, Japan, Switzerland, UK, and USA, along with a developed countries index (MXWO). The returns are total gross returns with dividends reinvested. For robustness, we also use the world index (MXWD) and look at the regular return index. We source data from Bloomberg and the MSCI. Data are end-of-period returns for weekly and monthly frequencies and span all US trading days from January 4th, 1999 to December 31st, 2018.

\subsection{Dataset 5: Size/Book-to-Market}

\noindent{}We employ returns on the 6 ($2\times3$) portfolios sorted by size and book-to-market. Remaining details (e.g., source, weighting, frequency, date range) are the same as in Section~\ref{subsec:ind}. 

\subsection{Dataset 6: Momentum Portfolios}

\noindent{}This dataset consists of returns on the 10 portfolios sorted by momentum. Remaining details (e.g., source, weighting, frequency, date range) are the same as in Section~\ref{subsec:ind}. 

\section{Assessment Criteria}\label{sec:assess}

\noindent{}We assess performance through three industry-standard metrics: 
Sharpe ratio, portfolio volatility, and Sharpe ratio net of turnover costs.\footnote{Covariance-based methods such as the minimum variance portfolio may lower variance relative to the 1/N portfolio and thus raise Sharpe ratios. We therefore also consider returns. While the naive strategy performs well on dataset 1, and despite with weaker dominance over the naive strategy on dataset 6, other strategies dominate the naive strategy on datasets 2 through 5. Results are available upon request.} For each metric, we estimate the statistical significance of the difference in the estimated metric from that of the 1/N strategy.

\subsection{Sharpe Ratio}

\noindent{}The Sharpe ratio measures reward to risk from a portfolio strategy, i.e., expected return per standard deviation. To test for differences between the Sharpe ratio from investing according to the naive strategy and the Sharpe ratio from the strategy in question, we employ the robust inference methods of~\citet{ledoit2008robust}.


\subsection{Sharpe Ratio Adjusted for Turnover Cost}

\noindent{}We assume a proportional turnover cost of 0.5\% 
and calculate the expected returns net of the cost of rebalancing similar to~\citet{demiguel2014stock}
\begin{equation*}
\mathbf{r}_{t+1} = \left(1-\kappa\sum_{i=1}^N{|w_{i,t}-w_{i,(t-1)*}|}\right)(\mathbf{w}_t)'\mathbf{r_{t+1}},
\end{equation*}
where $w_{i,(t-1)*}^k$ is the weight in investment $i$ and time $t$ prior to rebalancing, $w_{i,t}$ is the weight suggested by the strategy at time $t$, i.e., after rebalancing, $\kappa$ is the proportional transaction cost, $\mathbf{w}_t$ is the vector of weights, and $\mathbf{r}_{t+1}$ is the return vector.\footnote{Several papers in the literature consider transaction costs of 10 or 50 basis points~(\citealp{kirby2012s}; \citealp{demiguel2014stock}) and others consider transactions costs that vary across stock size and through time~\citep{brandt2009parametric}. With high turnover, assuming 50 basis points 
transactions costs conservatively biases our models away from beating the 1/N strategy.} Rebalancing may occur each period. We compare the difference in Sharpe ratios between the expected net returns following both the specific strategy and the naive strategy. 

\subsection{Portfolio Volatility}

\noindent{}Assuming that the investor's goal is to minimize portfolio volatility, by analogy with~\citet{wang2015hedging}, we examine ranking portfolio strategies by out-of-sample volatility of returns. To be precise, we conduct the Brown-Forsythe F* test of unequal group variances. We also apply the Diebold-Mariano test in comparing forecast errors from a naive strategy with the strategy under consideration~\citep{diebold1995comparing}.\footnote{The forecast error is defined as the difference between expected returns using estimated portfolio weights and mean returns. The loss differential underlying the test looks at the difference of the squared forecast errors, and we calculate the 
the loss differential correcting for autocorrelation.} 
 The procedures allow testing whether the strategy is significantly more or less volatile relative to the naive strategy. 

\section{Empirical Results}\label{sec:results}


\noindent{}Our evidence suggests that the out-of-sample performance of portfolios whose only inputs are volatility estimates often weakly dominate that of the naive diversification portfolio. The minimum-variance, constrained minimum-variance, volatility-timing, and tangency portfolios have equivalent or superior Sharpe ratios, portfolio volatility, and Sharpe ratios adjusted for turnover costs relative to the naive portfolio in datasets 2, 3, 5, and 6, regardless of the econometric model used to estimate volatility. If we average the volatility estimates of all thirteen econometric models, we continue to obtain similar results. Our portfolio strategies perform well relative to the naive when applied to country stock-market indices. Our results are robust to value- and equal-weighting and to weekly and monthly frequency estimation.\footnote{We report only results for value-weighted data at weekly frequency.} We thus show that controlling for volatility in portfolio strategies delivers better performance than the naive portfolio.



In the next two subsections, we evaluate the performance of the econometric models. Specifically, we select an individual performance metric (e.g., the Sharpe ratio) and attempt to rank econometric models by performance consistency across datasets within a given portfolio strategy (e.g., minimum-variance). Ranking allows us to observe how models perform in an absolute sense across datasets. Often, however, one model has the highest Sharpe ratio yet the highest portfolio volatility. In the third subsection, we therefore undertake a holistic analysis that incorporates all three performance metrics to rank econometric models and portfolio strategies that consistently outperform the naive strategy.

\subsection{Sharpe Ratio}\label{subsec:sharpe}

\noindent{}We first evaluate the Sharpe ratio performance metric. In Tables~\ref{tab:mvsr}--\ref{tab:tansr}, we provide the Sharpe ratios associated with each of our thirteen econometric models when used as the input in each of the four portfolio strategies. Sharpe ratios are assessed across all six datasets with value-weighting at weekly frequency. We empirically test the difference in Sharpe ratios between each of the econometric models relative to the naive strategy and report significance levels. 
The row ordering reflects our attempt to rank the econometric models according to consistency of performance relative to the naive benchmark.

In the minimum-variance and constrained minimum-variance portfolios, the combined parameter (CP) model yields Sharpe ratios that are consistently and significantly higher than those of the naive benchmark. In fact, nearly all econometric models achieve significantly higher Sharpe ratios relative to the naive rule in datasets 2, 4, 5, and 6, while displaying broadly equivalent Sharpe ratios in datasets 1 and 3.\footnote{To explain the poorer performance of datasets 1 and 3, first, the literature consistently finds weak performance with the Fama-French dataset~\citep{demiguel2009optimal}; second, a simple correlation matrix of the six datasets shows that dataset 3 is the only dataset to be negatively correlated with the other datasets.} Consequentially, most econometric models, when combined with either the constrained or unconstrained minimum-variance portfolio, weakly dominate the naive benchmark. The few exceptions, which occur only in dataset 1, are the vector autoregression (VAR) and vector error-correction (VEC) models in the minimum-variance portfolio, and the regime-switching vector autoregression (RSVAR) model in both the constrained and unconstrained minimum-variance portfolios. Even our lowest ranked econometric model, the exponentially-weighted moving-average (EWMA) model, still weakly dominates the naive portfolio.

In the volatility-timing and tangency portfolios, the realized covariance (RCOV) model delivers Sharpe ratios that are consistently and significantly higher than those of the naive benchmark. Let us first examine the volatility-timing portfolio. Most econometric models achieve significantly higher Sharpe ratios relative to the naive rule in datasets 2, 4, 5, and 6, and similar Sharpe ratios in dataset 3. The only weakness again lies in dataset 1, where all models except RCOV underperform the naive rule. Turning our attention to the tangency portfolio, we observe significantly higher Sharpe ratios relative to the naive benchmark across models in datasets 5 and 6, and similar Sharpe ratios in the rest. Thus, we conclude the tangency portfolio weakly dominates the naive portfolio in terms of Sharpe ratio. As with the minimum-variance portfolios, the exponentially-weighted moving-average (EWMA) achieves the worst Sharpe ratios relative to the other econometric models yet still performs well in comparison to the naive strategy. 

Tables 
S1--S2 in 
the 
\ifblankver
online appendix 
\else
\href{http://www.michael-curran.com/research/volatility_appendix.pdf}{\textcolor{blue}{online appendix}} 
\fi
show that the results are robust to adjusting the Sharpe ratios for turnover costs.\footnote{Allocations can shift, requiring rebalancing turnover even for the naive portfolio. With turnover, expected returns are no larger, but standard deviations may be smaller or larger.} The only difference is that the BEKK- and ABEKK-GARCH models achieve the highest and most consistent turnover-cost-adjusted Sharpe ratio with the minimum-variance portfolio relative to the naive benchmark. Moreover, we determine results are robust to both equal-weighting and monthly frequency.  

\begin{figure}
\centering
\captionsetup{font=small,skip=0pt}
\caption{Sharpe Ratio Percentage Difference Relative to Naive}
\label{fig:sharpesubplot}
\begin{subfigure}{\textwidth}
\centering
\includegraphics[height=15.5cm,width=\columnwidth]{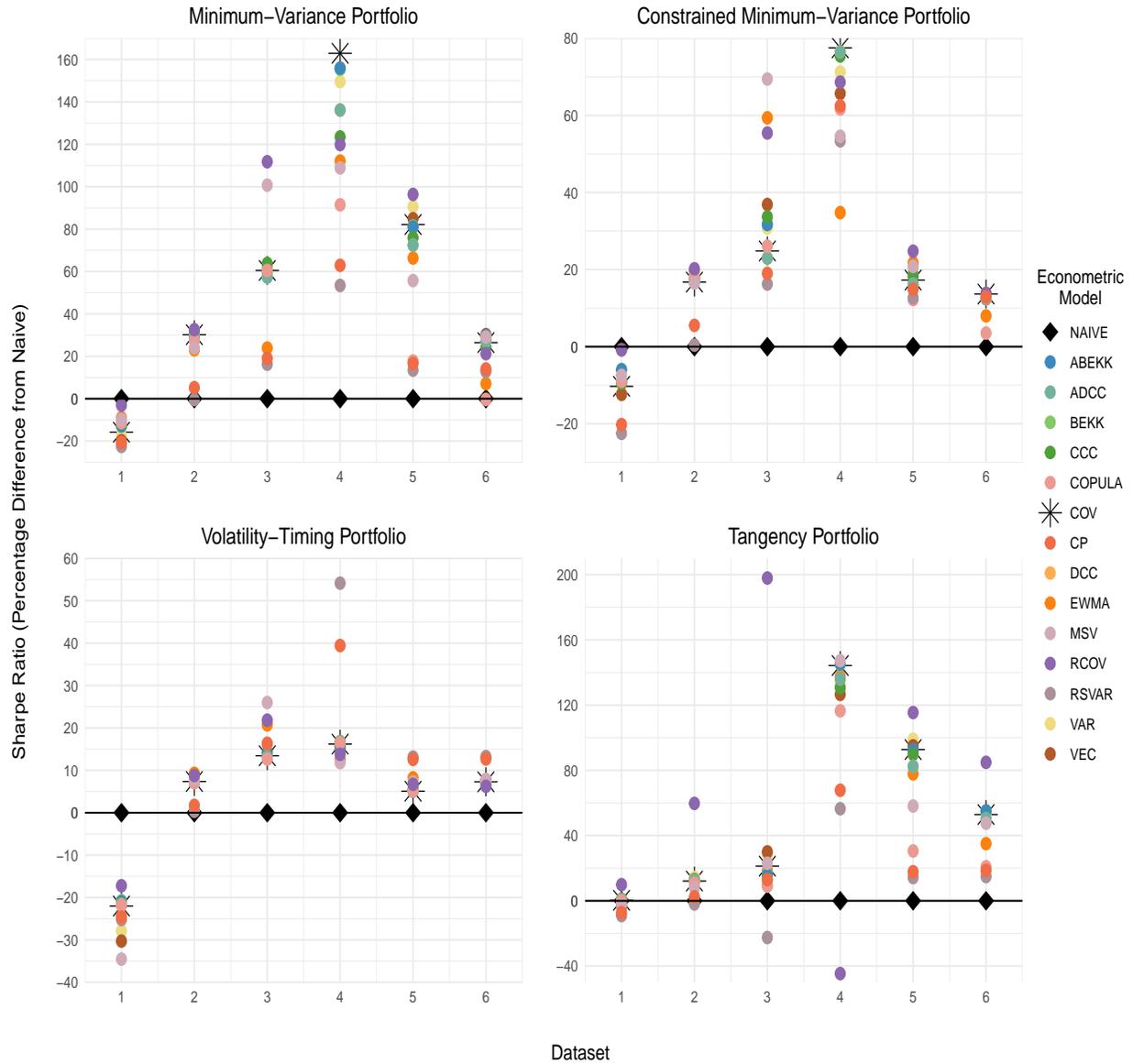}
\end{subfigure}
\floatfoot{\small Notes: See Section~\ref{sec:metrics} for explanations of econometric model abbreviations; CP denotes combined parameter model~\eqref{eq:comv}. Dataset 1: Fama-French portfolios; Dataset 2: industry portfolios; Dataset 3: sector portfolios; Dataset 4: international equity indices; Dataset 5: portfolios sorted by size/book-to-market; Dataset 6: momentum portfolios. 
Results are for value-weighted data at weekly frequency. Note that standard errors of the Sharpe ratio for the tangency portfolio are mostly large, rendering most Sharpe ratios for the tangency portfolio statistically insignificantly different to Sharpe ratios for the naive portfolio.}
\end{figure}

The Sharpe ratios of our portfolio strategies relative to the naive strategy are not just statistically significant but \textit{economically} significant. Figure~\ref{fig:sharpesubplot} illustrates this point. On average, our portfolio strategies achieve Sharpe ratios that are 30\% higher than the naive, across all six datasets, punctuated by the minimum-variance portfolio at 47\%. We obtain a similar message in Figure 
S1.1 from the 
\ifblankver
online appendix 
\else
\href{http://www.michael-curran.com/research/volatility_appendix.pdf}{\textcolor{blue}{online appendix}} 
\fi
when we adjust 
for turnover costs.

\subsection{Portfolio Volatility}\label{subsec:vol}

\noindent{}
Tables~\ref{tab:mvpv}--\ref{tab:tanpv} 
provide the standard deviations of the returns associated with each of our thirteen econometric models when used as the input in each of the four portfolio strategies. 
The row ordering ranks the econometric models according to consistency of performance relative to the naive benchmark.


In the minimum-variance and constrained minimum-variance portfolios, the exponentially-weighted moving-average (EWMA), realized covariance (RCOV), and multivariate stochastic volatility (MSV) models exhibit significantly lower portfolio volatility relative to the naive portfolio across all datasets. More importantly, most econometric models \textit{strictly} dominate the naive portfolio in terms of volatility performance. The two exceptions, regime-switching vector autoregression (RSVAR) and combined parameter (CP), which happen to be the worst ranking models, still \textit{weakly} dominate the naive benchmark.

For the volatility-timing portfolio, the EWMA model delivers the best results across datasets. Moreover, every econometric model weakly dominates the naive benchmark. For the tangency portfolio, the COPULA achieves the lowest portfolio volatility across datasets. All econometric models, except RCOV and VEC in dataset 4, weakly dominate the naive benchmark. In addition, the MSV and RCOV models are consistent runner-ups in both the volatility-timing and tangency portfolio strategies. Although the RSVAR and CP models yield the highest volatility, both still weakly dominate the naive portfolio.



\begin{figure}
\centering
\captionsetup{font=small,skip=0pt}
\caption{Portfolio Volatility Percentage Difference Relative to Naive}
\label{fig:volsubplot}
\begin{subfigure}{\textwidth}
\centering
\includegraphics[height=15.5cm,width=\columnwidth]{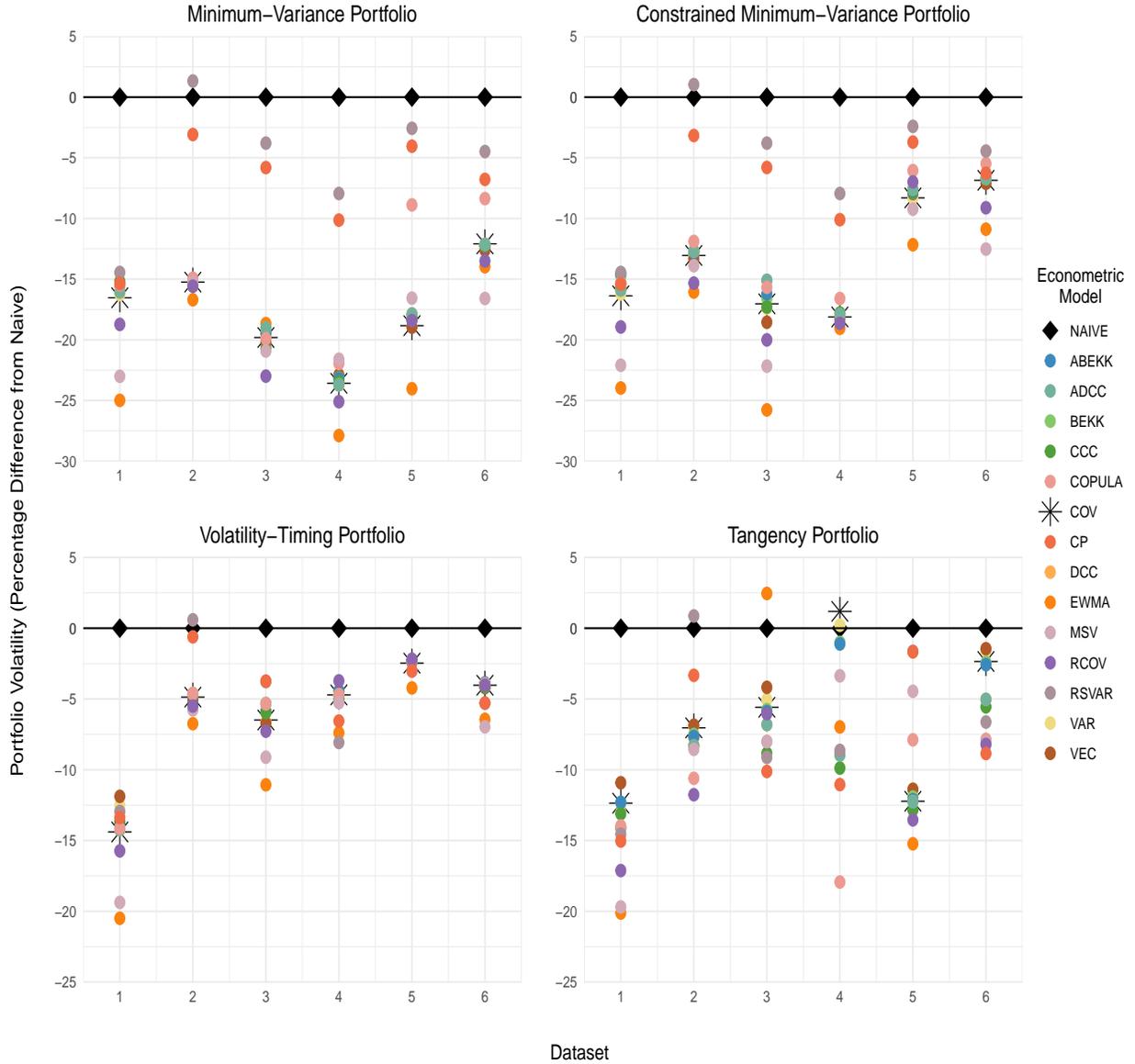}
\end{subfigure}
\floatfoot{\small Notes: See Section~\ref{sec:metrics} for explanations of econometric model abbreviations; CP denotes combined parameter model~\eqref{eq:comv}. Dataset 1: Fama-French portfolios; Dataset 2: industry portfolios; Dataset 3: sector portfolios; Dataset 4: international equity indices; Dataset 5: portfolios sorted by size/book-to-market; Dataset 6: momentum portfolios.
Results are for value-weighted data at weekly frequency.}
\end{figure}

The volatility of our portfolio strategies relative to the naive strategy is \textit{economically} significant. Figure~\ref{fig:volsubplot} illustrates this point. On average, our portfolio strategies are 9\% less volatile, across all six datasets, with the minimum-variance portfolio at 10\% lower volatility. 

\clearpage

\subsection{Portfolio Strategies}\label{subsec:pm}

\noindent{}We undertake a holistic evaluation of the econometric models. 
In Tables~\ref{tab:cov}--\ref{tab:msv}, for each econometric model, we compare 
the Sharpe ratio, Sharpe ratio adjusted for turnover cost, and portfolio volatility of each portfolio strategy across all six datasets. 
All of our econometric models are broadly successful at outperforming the naive portfolio; that is, they achieve higher Sharpe ratios, lower turnover costs, lower portfolio volatility, or some combination. The few exceptions are concentrated where the volatility-timing portfolio is applied to dataset 1.  
To identify the best performers, we develop a simple heuristic to score individual econometric models: we sum the instances where the model outperforms the naive benchmark and subtract the instances where the model underperforms.\footnote{To clarify, ``$\checkmark$" $= 1$, ``$\checkmark$*" $= 2/3$, `` " (blanks) $= 0$, and ``$\times$" $= -1.$ We discount results that are significant at the 10\% level by assigning a value of only 2/3 instead of 1. }

The multivariate GARCH models achieve the highest scores relative to other econometric models when applied to the minimum-variance and constrained minimum-variance strategies. With GARCH estimates of the covariance matrix, these portfolio strategies weakly dominate the naive benchmark. The constant conditional correlation (CCC) performs especially well without short sales. For the volatility-timing and tangency portfolios, the realized covariance (RCOV) model exhibits the most impressive results relative to the naive rule. Specifically, RCOV weakly dominates the naive rule across every dataset when paired with the volatility-timing strategy, and across five out of six datasets when paired with the tangency portfolio. In general, our results suggest the multivariate GARCH and RCOV models are better alternatives to the often-used sample covariance (COV) matrix for portfolio construction. The sample covariance matrix (COV) performs at the median relative to the other econometric models. 
While the convenience of COV makes it attractive to researchers, our analysis shows there are returns to using more sophisticated methods to forecast volatility. The worst-ranking econometric models are the regime-switching vector autoregression (RSVAR) and exponentially-weighted moving-average (EWMA) models. Nonetheless, both of these models perform at least as well as the naive benchmark in every dataset except the Fama-French 3-factor.

As a final assessment, we naively average the estimated conditional volatilities of all thirteen econometric models to form a combined parameter (CP) model; 
 see Table~\ref{tab:msv}. 
The main takeaway from this exercise is that controlling for volatility in a portfolio delivers performance metrics that are generally at least as strong as the naive strategy.

\section{Conclusion}\label{sec:conc}

\noindent{}We evaluate the out-of-sample performance of mean-variance strategies relying solely upon the second moment relative to the naive benchmark. 
Using fourteen econometric models across six datasets at weekly frequency, we show that the minimum-variance, constrained minimum-variance, and volatility-timing strategies generally achieve higher Sharpe ratios, lower turnover costs, and lower portfolio volatility that are \textit{economically} significant relative to naive diversification. Whenever mean-variance strategies do not significantly outperform the naive rule, they usually match and only rarely lose to it. 

We identify the econometric models that most consistently and significantly outperform the 1/N benchmark. First, we show that the multivariate GARCH models weakly dominate the naive rule when applied to the minimum-variance and constrained minimum-variance strategies. Next, we demonstrate that the realized covariance model achieves impressive results when paired with the volatility-timing and tangency portfolios. Even our ``worst-performing" econometric models still perform at least as well as the naive rule in all but one dataset. Third, we illustrate that if one prioritizes the Sharpe ratio, then the combined parameter and realized covariance models are excellent choices, even after controlling for turnover costs. Finally, we show the exponentially-weighted moving-average and multivariate stochastic volatility models consistently deliver low portfolio volatility. 


With the difficulty in consistently outperforming the strategy, the 1/N naive diversification should serve as a benchmark for practitioners and academics. We empirically demonstrate that an important source of the naive portfolio puzzle is the quality of the econometric volatility inputs to the mean-variance portfolio strategies. With improved estimates, mean-variance models can beat the naive portfolio strategy. 
Our findings imply that improving the estimation of return moments should be prioritized. 

\begin{singlespacing}
\bibliographystyle{ecca}
\bibliography{myrefs}
\end{singlespacing}


\newpage

\begin{table}[htbp!]
\captionsetup{justification=centering,font=small,skip=0pt}
\begin{center}
\scalebox{0.9}{
\begin{tabular}{@{}ld{4.6}d{4.6}d{4.6}d{4.6}d{4.6}d{4.6}@{}}
\toprule
       &\multicolumn{1}{c}{ \text{Dataset 1}} &\multicolumn{1}{c}{ \text{Dataset 2}} &\multicolumn{1}{c}{ \text{Dataset 3}} &\multicolumn{1}{c}{ \text{Dataset 4}} &\multicolumn{1}{c}{ \text{Dataset 5}} &\multicolumn{1}{c}{ \text{Dataset 6}} \\
\midrule
\textit{Naive}  & 0.112     & 0.101     & 0.035     & 0.058     & 0.100     & 0.087     \\ 
\midrule
\textit{Minimum}-&&&&&&\\
\textit{Variance}&&&&&&\\
CP     & 0.089     & 0.106${**}$     & 0.042     & 0.095${**}$     & 0.117${***}$     & 0.099${***}$     \\
ABEKK  & 0.098     & 0.131${**}$     & 0.056     & 0.149${**}$     & 0.181${***}$     & 0.113${**}$     \\
BEKK   & 0.098     & 0.131${**}$     & 0.057     & 0.148${**}$     & 0.181${***}$     & 0.113${**}$     \\
DCC    & 0.100     & 0.131${**}$     & 0.055     & 0.137${**}$     & 0.172${***}$     & 0.110${**}$     \\
ADCC   & 0.100     & 0.131${**}$     & 0.055     & 0.137${**}$     & 0.172${***}$     & 0.110${**}$     \\
CCC    & 0.099     & 0.131${**}$     & 0.057     & 0.130${**}$     & 0.176${***}$     & 0.108${**}$     \\
VAR    & (0.093${*}$)     & 0.134${**}$     & 0.057     & 0.145${**}$     & 0.191${***}$     & 0.110${**}$     \\
VEC    & (0.090${*}$)     & 0.132${**}$     & 0.056     & 0.149${**}$     & 0.185${***}$     & 0.113${**}$     \\
COV    & 0.095     & 0.132${**}$     & 0.056     & 0.153${**}$     & 0.182${***}$     & 0.110${*}$     \\
RSVAR  & (0.087${*}$)     & 0.101     & 0.041     & 0.089${**}$     & 0.113${***}$     & 0.098${***}$     \\
RCOV   & 0.109     & 0.134${*}$     & 0.074     & 0.128${**}$     & 0.196${***}$     & 0.105     \\
COPULA & 0.100     & 0.129${**}$     & 0.056     & 0.111${*}$     & 0.118${***}$     & 0.086     \\
EWMA   & 0.102     & 0.124     & 0.043     & 0.123${*}$     & 0.166${***}$     & 0.093     \\
MSV    & 0.101     & 0.125     & 0.070     & 0.121${*}$     & 0.156${***}$     & 0.112     \\
\midrule
\textit{Constrained}&&&&&&\\
\textit{Minimum}-&&&&&&\\
\textit{Variance}&&&&&&\\
CCC    & 0.102     & 0.119${**}$     & 0.047     & 0.102${**}$     & 0.118${***}$     & 0.098${***}$     \\
CP     & 0.090     & 0.107${**}$     & 0.042     & 0.094${**}$     & 0.115${***}$     & 0.098${***}$     \\
COV    & 0.101     & 0.118${*}$     & 0.044     & 0.103${***}$     & 0.117${***}$     & 0.099${***}$     \\
ABEKK  & 0.106     & 0.118${*}$     & 0.046     & 0.102${***}$     & 0.119${***}$     & 0.098${***}$     \\
ADCC   & 0.102     & 0.119${*}$     & 0.043     & 0.102${***}$     & 0.116${***}$     & 0.098${***}$     \\
VAR    & 0.099     & 0.118${*}$     & 0.046     & 0.099${***}$     & 0.120${***}$     & 0.098${**}$     \\
VEC    & 0.098     & 0.119${*}$     & 0.048     & 0.096${**}$     & 0.116${***}$     & 0.099${***}$     \\
DCC    & 0.102     & 0.119${*}$     & 0.043     & 0.102${**}$     & 0.116${***}$     & 0.098${**}$     \\
MSV    & 0.104     & 0.118${*}$     & 0.059     & 0.090${**}$     & 0.121${***}$     & 0.098${*}$     \\
RCOV   & 0.111     & 0.121     & 0.055     & 0.098${**}$     & 0.125${***}$     & 0.099${**}$     \\
BEKK   & 0.106     & 0.118     & 0.046     & 0.102${**}$     & 0.118${***}$     & 0.098${**}$     \\
COPULA & 0.102     & 0.120${**}$     & 0.044     & 0.094${**}$     & 0.112${***}$     & 0.090     \\
RSVAR  & (0.087${*}$)     & 0.101     & 0.041     & 0.089${**}$     & 0.113${***}$     & 0.098${***}$     \\
EWMA   & 0.111     & 0.121${**}$     & 0.056     & 0.078     & 0.122${***}$     & 0.094     \\
\bottomrule
\end{tabular}
}
\end{center}
\small {
\begin{flushleft}
\begin{minipage}{16.2cm}
\emph{Notes:} See 
Section~\ref{sec:metrics} for explanations of econometric model abbreviations; CP denotes combined parameter model~\eqref{eq:comv}. Dataset 1: Fama-French portfolios; Dataset 2: industry portfolios; Dataset 3: sector portfolios; Dataset 4: international equity indices; Dataset 5: portfolios sorted by size/book-to-market; Dataset 6: momentum portfolios. Results are for value-weighted data at weekly frequency. * significant at 10\%; ** significant at 5\%; *** significant at 1\%. Significance corresponds to the~\citet{ledoit2008robust} robust test for differences between the Sharpe ratio and that of the naive strategy. Numbers in parentheses are statistically significantly worse than those of the naive strategy.
\end{minipage}
\end{flushleft}
}
\caption{Sharpe Ratio: Minimum-Variance and Constrained Minimum-Variance}
\label{tab:mvsr}
\end{table}

\begin{table}[htbp!]
\captionsetup{justification=centering,font=small,skip=0pt}
\begin{center}
\scalebox{0.9}{
\begin{tabular}{@{}ld{4.6}d{4.6}d{4.6}d{4.6}d{4.6}d{4.6}@{}}
\toprule
       &\multicolumn{1}{c}{ \text{Dataset 1}} &\multicolumn{1}{c}{ \text{Dataset 2}} &\multicolumn{1}{c}{ \text{Dataset 3}} &\multicolumn{1}{c}{ \text{Dataset 4}} &\multicolumn{1}{c}{ \text{Dataset 5}} &\multicolumn{1}{c}{ \text{Dataset 6}} \\
\midrule       
\textit{Naive}  & 0.112     & 0.101     & 0.035     & 0.058     & 0.100     & 0.087     \\ 
\midrule       
\textit{Volatility}&&&&&&\\
\textit{Timing}&&&&&&\\
RCOV   & 0.093     & 0.110${***}$     & 0.043${*}$     & 0.066${***}$     & 0.107${***}$     & 0.092${***}$     \\
CCC    & (0.088${*}$)     & 0.108${***}$     & 0.040     & 0.068${***}$     & 0.106${***}$     & 0.093${***}$     \\
BEKK   & (0.089${*}$)     & 0.109${***}$     & 0.040     & 0.067${***}$     & 0.105${***}$     & 0.093${***}$     \\
ABEKK  & (0.089${*}$)     & 0.109${***}$     & 0.040     & 0.067${***}$     & 0.105${***}$     & 0.093${***}$     \\
COPULA & (0.088${*}$)     & 0.108${***}$     & 0.040     & 0.068${***}$     & 0.105${***}$     & 0.093${***}$     \\
ADCC   & (0.088${*}$)     & 0.108${***}$     & 0.040     & 0.068${***}$     & 0.105${***}$     & 0.093${***}$     \\
DCC    & (0.088${*}$)     & 0.108${**}$     & 0.040     & 0.068${***}$     & 0.105${***}$     & 0.093${***}$     \\
COV    & (0.088${**}$)     & 0.108${***}$     & 0.040     & 0.067${***}$     & 0.105${***}$     & 0.093${***}$     \\
VAR    & (0.081${***}$)     & 0.109${***}$     & 0.040     & 0.067${**}$     & 0.106${***}$     & 0.093${***}$     \\
VEC    & (0.078${***}$)     & 0.108${**}$     & 0.040     & 0.067${***}$     & 0.105${***}$     & 0.094${***}$     \\
MSV    & (0.074${**}$)     & 0.109${**}$     & 0.044${*}$     & 0.065${*}$     & 0.107${***}$     & 0.094${**}$     \\
RSVAR  & (0.084${*}$)     & 0.101     & 0.041     & 0.090${**}$     & 0.113${***}$     & 0.098${***}$     \\
CP     & (0.085${*}$)     & 0.103     & 0.041     & 0.081${**}$     & 0.113${***}$     & 0.098${***}$     \\
EWMA   & (0.085${*}$)     & 0.110${**}$     & 0.042     & 0.065     & 0.108${***}$     & 0.093${**}$     \\
\midrule
\textit{Tangency}&&&&&&\\
RCOV   & 0.124     & 0.161${***}$     & 0.104${**}$     & 0.032     & 0.215${***}$     & 0.160${***}$     \\
RSVAR  & 0.102     & 0.099     & 0.027     & 0.091${**}$     & 0.114${***}$     & 0.100${***}$     \\
CP     & 0.104     & 0.103     & 0.040     & 0.097${**}$     & 0.118${***}$     & 0.103${**}$     \\
DCC    & 0.113     & 0.114     & 0.038     & 0.137     & 0.182${***}$     & 0.130${***}$     \\
VAR    & 0.109     & 0.115     & 0.045     & 0.135     & 0.199${***}$     & 0.133${**}$     \\
VEC    & 0.108     & 0.113     & 0.046     & 0.132     & 0.195${***}$     & 0.135${**}$     \\
BEKK   & 0.113     & 0.113     & 0.041     & 0.142     & 0.192${***}$     & 0.134${**}$     \\
ABEKK  & 0.113     & 0.113     & 0.041     & 0.142     & 0.192${***}$     & 0.134${**}$     \\
COV    & 0.113     & 0.113     & 0.043     & 0.142     & 0.193${***}$     & 0.133${**}$     \\
CCC    & 0.113     & 0.114     & 0.040     & 0.134     & 0.190${***}$     & 0.130${**}$     \\
ADCC   & 0.113     & 0.114     & 0.038     & 0.137     & 0.182${***}$     & 0.130${**}$     \\
MSV    & 0.110     & 0.109     & 0.043     & 0.144${*}$     & 0.158${***}$     & 0.128${*}$     \\
EWMA   & 0.112     & 0.101     & 0.042     & 0.138     & 0.178${***}$     & 0.117     \\
COPULA & 0.112     & 0.112     & 0.038     & 0.126     & 0.131${***}$     & 0.105     \\
\bottomrule
\end{tabular}%
}
\end{center}
\small {
\begin{flushleft}
\begin{minipage}{16.2cm}
\emph{Notes:} See 
Section~\ref{sec:metrics} for explanations of econometric model abbreviations; CP denotes combined parameter model~\eqref{eq:comv}. Dataset 1: Fama-French portfolios; Dataset 2: industry portfolios; Dataset 3: sector portfolios; Dataset 4: international equity indices; Dataset 5: portfolios sorted by size/book-to-market; Dataset 6: momentum portfolios. Results are for value-weighted data at weekly frequency. * significant at 10\%; ** significant at 5\%; *** significant at 1\%. Significance corresponds to the~\citet{ledoit2008robust} robust test for differences between the Sharpe ratio and that of the naive strategy. Numbers in parentheses are statistically significantly worse than those of the naive strategy.
\end{minipage}
\end{flushleft}
}
\caption{Sharpe Ratio: Volatility-Timing and Tangency}
\label{tab:tansr}
\end{table}

\begin{table}[htbp!]
\captionsetup{justification=centering,font=small,skip=0pt}
\begin{center}
\scalebox{0.9}{
\begin{tabular}{@{}ld{4.6}d{4.6}d{4.6}d{4.6}d{4.6}d{4.6}@{}}
\toprule
       &\multicolumn{1}{c}{ \text{Dataset 1}} &\multicolumn{1}{c}{ \text{Dataset 2}} &\multicolumn{1}{c}{ \text{Dataset 3}} &\multicolumn{1}{c}{ \text{Dataset 4}} &\multicolumn{1}{c}{ \text{Dataset 5}} &\multicolumn{1}{c}{ \text{Dataset 6}} \\
\midrule
\textit{Naive}  & 0.002     & 0.005     & 0.024     & 0.024     & 0.005     & 0.005     \\ 
\midrule
\textit{Minimum}-&&&&&&\\
\textit{Variance}&&&&&&\\
EWMA   & 0.001${***}$     & 0.004${***}$     & 0.019${***}$     & 0.017${***}$     & 0.004${***}$     & 0.005${***}$     \\
MSV    & 0.001${***}$     & 0.004${***}$     & 0.019${***}$     & 0.019${***}$     & 0.004${***}$     & 0.004${***}$     \\
RCOV   & 0.002${***}$     & 0.004${***}$     & 0.018${***}$     & 0.018${***}$     & 0.004${***}$     & 0.005${***}$     \\
DCC    & 0.002${***}$     & 0.004${***}$     & 0.019${***}$     & 0.018${***}$     & 0.004${***}$     & 0.005${***}$     \\
ADCC   & 0.002${***}$     & 0.004${***}$     & 0.019${***}$     & 0.018${***}$     & 0.004${***}$     & 0.005${***}$     \\
VAR    & 0.002${***}$     & 0.004${***}$     & 0.019${***}$     & 0.018${***}$     & 0.004${***}$     & 0.005${***}$     \\
VEC    & 0.002${***}$     & 0.004${***}$     & 0.019${***}$     & 0.019${***}$     & 0.004${***}$     & 0.005${***}$     \\
BEKK   & 0.002${***}$     & 0.004${***}$     & 0.019${***}$     & 0.019${***}$     & 0.004${***}$     & 0.005${***}$     \\
ABEKK  & 0.002${***}$     & 0.004${***}$     & 0.019${***}$     & 0.019${***}$     & 0.004${***}$     & 0.005${***}$     \\
CCC    & 0.002${***}$     & 0.004${***}$     & 0.019${***}$     & 0.019${***}$     & 0.004${***}$     & 0.005${***}$     \\
COV    & 0.002${***}$     & 0.004${***}$     & 0.019${***}$     & 0.019${***}$     & 0.004${***}$     & 0.005${***}$     \\
COPULA & 0.002${***}$     & 0.004${***}$     & 0.019${***}$     & 0.019${***}$     & 0.005${***}$     & 0.005${***}$     \\
CP     & 0.002${***}$     & 0.005     & 0.022     & 0.022${**}$     & 0.005${*}$     & 0.005${**}$     \\
RSVAR  & 0.002${***}$     & 0.005     & 0.023     & 0.022${*}$     & 0.005     & 0.005     \\
\midrule
\textit{Constrained}&&&&&&\\
\textit{Minimum}-&&&&&&\\
\textit{Variance}&&&&&&\\
EWMA   & 0.001${***}$     & 0.004${***}$     & 0.017${***}$     & 0.020${***}$     & 0.004${***}$     & 0.005${***}$     \\
MSV    & 0.001${***}$     & 0.004${***}$     & 0.018${***}$     & 0.020${***}$     & 0.005${***}$     & 0.005${***}$     \\
RCOV   & 0.002${***}$     & 0.004${***}$     & 0.019${***}$     & 0.020${***}$     & 0.005${***}$     & 0.005${***}$     \\
VEC    & 0.002${***}$     & 0.004${***}$     & 0.019${***}$     & 0.020${***}$     & 0.005${***}$     & 0.005${**}$     \\
CCC    & 0.002${***}$     & 0.004${***}$     & 0.019${***}$     & 0.020${***}$     & 0.005${***}$     & 0.005${**}$     \\
VAR    & 0.002${***}$     & 0.004${***}$     & 0.019${**}$     & 0.020${***}$     & 0.005${***}$     & 0.005${**}$     \\
BEKK   & 0.002${***}$     & 0.004${***}$     & 0.020${**}$     & 0.020${***}$     & 0.005${***}$     & 0.005${**}$     \\
ABEKK  & 0.002${***}$     & 0.004${***}$     & 0.020${**}$     & 0.020${***}$     & 0.005${***}$     & 0.005${**}$     \\
DCC    & 0.002${***}$     & 0.004${***}$     & 0.020${**}$     & 0.020${***}$     & 0.005${***}$     & 0.005${**}$     \\
ADCC   & 0.002${***}$     & 0.004${***}$     & 0.020${**}$     & 0.020${***}$     & 0.005${***}$     & 0.005${**}$     \\
COV    & 0.002${***}$     & 0.004${***}$     & 0.020${**}$     & 0.020${***}$     & 0.005${***}$     & 0.005${**}$     \\
COPULA & 0.002${***}$     & 0.004${***}$     & 0.020${**}$     & 0.020${***}$     & 0.005${**}$     & 0.005${*}$     \\
CP     & 0.002${***}$     & 0.005     & 0.022     & 0.022${**}$     & 0.005${*}$     & 0.005${*}$     \\
RSVAR  & 0.002${***}$     & 0.005     & 0.023     & 0.022${*}$     & 0.005     & 0.005     \\
\bottomrule
\end{tabular}%
}
\end{center}
\small {
\begin{flushleft}
\begin{minipage}{16.2cm}
\emph{Notes:} See 
Section~\ref{sec:metrics} for explanations of econometric model abbreviations; CP denotes combined parameter model~\eqref{eq:comv}. Dataset 1: Fama-French portfolios; Dataset 2: industry portfolios; Dataset 3: sector portfolios; Dataset 4: international equity indices; Dataset 5: portfolios sorted by size/book-to-market; Dataset 6: momentum portfolios. Results are for value-weighted data at weekly frequency. * significant at 10\%; ** significant at 5\%; *** significant at 1\%. Significance corresponds to the Brown-Forsythe F* test for unequal group variances. Results from the Diebold-Mariano~\citep{diebold1995comparing} test for differences in variance relative to the naive portfolio are similar.
\end{minipage}
\end{flushleft}
}
\caption{Portfolio Volatility: Minimum-Variance and Constrained Minimum-Variance}
\label{tab:mvpv}
\end{table}

\begin{table}[htbp!]
\captionsetup{justification=centering,font=small,skip=0pt}
\begin{center}
\scalebox{0.9}{
\begin{tabular}{@{}ld{4.6}d{4.6}d{4.6}d{4.6}d{4.6}d{4.6}@{}}
\toprule
       &\multicolumn{1}{c}{ \text{Dataset 1}} &\multicolumn{1}{c}{ \text{Dataset 2}} &\multicolumn{1}{c}{ \text{Dataset 3}} &\multicolumn{1}{c}{ \text{Dataset 4}} &\multicolumn{1}{c}{ \text{Dataset 5}} &\multicolumn{1}{c}{ \text{Dataset 6}} \\
\midrule       
\textit{Naive}  & 0.002     & 0.005     & 0.024     & 0.024     & 0.005     & 0.005     \\ 
\midrule
\textit{Volatility}&&&&&&\\
\textit{Timing}&&&&&&\\
EWMA   & 0.001${***}$     & 0.004${***}$    & 0.021${*}$     & 0.022     & 0.005     & 0.005${**}$     \\
MSV    & 0.002${***}$     & 0.004${**}$     & 0.021     & 0.023     & 0.005     & 0.005${**}$     \\
RCOV   & 0.002${***}$     & 0.004${**}$     & 0.022     & 0.023     & 0.005     & 0.005     \\
COPULA & 0.002${***}$     & 0.005${*}$     & 0.022     & 0.023     & 0.005     & 0.005     \\
BEKK   & 0.002${***}$     & 0.005${*}$     & 0.022     & 0.023     & 0.005     & 0.005     \\
ABEKK  & 0.002${***}$     & 0.005${*}$     & 0.022     & 0.023     & 0.005     & 0.005     \\
CCC    & 0.002${***}$     & 0.005${*}$     & 0.022     & 0.023     & 0.005     & 0.005     \\
DCC    & 0.002${***}$     & 0.005${*}$     & 0.022     & 0.023     & 0.005     & 0.005     \\
ADCC   & 0.002${***}$     & 0.005${*}$     & 0.022     & 0.023     & 0.005     & 0.005     \\
COV    & 0.002${***}$     & 0.005${*}$     & 0.022     & 0.023     & 0.005     & 0.005     \\
RSVAR  & 0.002${**}$     & 0.005     & 0.023     & 0.022${*}$     & 0.005     & 0.005${*}$     \\
VAR    & 0.002${**}$     & 0.005${*}$     & 0.022     & 0.023     & 0.005     & 0.005     \\
VEC    & 0.002${**}$     & 0.005${*}$     & 0.022     & 0.023     & 0.005     & 0.005     \\
CP     & 0.002${***}$     & 0.005     & 0.023     & 0.023     & 0.005     & 0.005     \\
\midrule
\textit{Tangency}&&&&&&\\
COPULA & 0.002${***}$     & 0.004${***}$     & 0.022     & 0.020${***}$     & 0.005${***}$     & 0.005${**}$     \\
MSV    & 0.002${***}$     & 0.004${***}$     & 0.022     & 0.023     & 0.005${***}$     & 0.005${**}$     \\
RCOV   & 0.002${***}$     & 0.004${***}$     & 0.022     & (0.025${*}$)  
& 0.004${***}$     & 0.005${**}$    \\
VAR    & 0.002${***}$     & 0.004${***}$     & 0.022     & 0.024     & 0.004${***}$     & 0.005     \\
BEKK   & 0.002${***}$     & 0.004${***}$     & 0.022     & 0.024     & 0.004${***}$     & 0.005     \\
ABEKK  & 0.002${***}$     & 0.004${***}$     & 0.022     & 0.024     & 0.004${***}$     & 0.005     \\
CCC    & 0.002${***}$     & 0.004${***}$     & 0.021     & 0.022     & 0.004${***}$     & 0.005     \\
DCC    & 0.002${***}$     & 0.004${***}$     & 0.022     & 0.022     & 0.004${***}$     & 0.005     \\
ADCC   & 0.002${***}$     & 0.004${***}$     & 0.022     & 0.022     & 0.004${***}$     & 0.005     \\
COV    & 0.002${***}$     & 0.004${***}$     & 0.022     & 0.025     & 0.004${***}$     & 0.005     \\
EWMA   & 0.001${***}$     & 0.004${***}$     & 0.024     & 0.023     & 0.004${***}$     & 0.005     \\
VEC    & 0.002${**}$     & 0.004${***}$     & 0.023     & (0.026${*}$)     & 0.004${***}$     & 0.005     \\
CP     & 0.002${***}$     & 0.005     & 0.021     & 0.022${**}$     & 0.005     & 0.005${**}$     \\
RSVAR  & 0.002${***}$     & 0.005     & 0.021     & 0.022${*}$     & 0.005     & 0.005     \\
\bottomrule
\end{tabular}%
}
\end{center}
\small {
\begin{flushleft}
\begin{minipage}{16.2cm}
\emph{Notes:} See 
Section~\ref{sec:metrics} for explanations of econometric model abbreviations; CP denotes combined parameter model~\eqref{eq:comv}. Dataset 1: Fama-French portfolios; Dataset 2: industry portfolios; Dataset 3: sector portfolios; Dataset 4: international equity indices; Dataset 5: portfolios sorted by size/book-to-market; Dataset 6: momentum portfolios. Results are for value-weighted data at weekly frequency. * significant at 10\%; ** significant at 5\%; *** significant at 1\%. Significance corresponds to the Brown-Forsythe F* test for unequal group variances. Results from the Diebold-Mariano~\citep{diebold1995comparing} test for differences in variance relative to the naive portfolio are similar.
\end{minipage}
\end{flushleft}
}
\caption{Portfolio Volatility: Volatility-Timing and Tangency}
\label{tab:tanpv}
\end{table}

\begin{table}[htbp!]
\captionsetup{justification=centering,font=small,skip=0pt}
\scalebox{0.82}{
\begin{tabular}{c}
COV \\
\end{tabular}
}\\
\scalebox{0.87}{
\begin{subtable}[t]{0.55\textwidth}
\centering
\begin{tabular}{@{}lllllll@{}}
\toprule
MVP & 1 & 2 & 3 & 4 & 5 & 6 \\ \midrule
SR Net of Turnover &  & $\checkmark$ &  & $\checkmark$ & $\checkmark$ & $\checkmark$* \\
Sharpe Ratio &  & $\checkmark$ &  & $\checkmark$ & $\checkmark$ & $\checkmark$* \\
Portfolio Volatility & $\checkmark$ & $\checkmark$ & $\checkmark$ & $\checkmark$ &$\checkmark$  &$\checkmark$  \\ 
\midrule
VT  & 1 & 2  & 3 & 4   & 5 & 6 \\ \midrule
SR Net of Turnover & $\times$ & $\checkmark$ &  & $\checkmark$ & $\checkmark$ & $\checkmark$ \\
Sharpe Ratio  & $\times$ & $\checkmark$ &  & $\checkmark$ & $\checkmark$ & $\checkmark$ \\
Portfolio Volatility & $\checkmark$ & $\checkmark$* &  &  &   &   \\ \bottomrule
\end{tabular}%
\end{subtable}
\hfill
\begin{subtable}[t]{0.55\textwidth}
\centering
\begin{tabular}{@{}lllllll@{}}
\toprule
con-MVP & 1 & 2 & 3 & 4 & 5 & 6 \\ \midrule
SR Net of Turnover &  & $\checkmark$* &  & $\checkmark$ & $\checkmark$ & $\checkmark$ \\
Sharpe Ratio &  & $\checkmark$* &  & $\checkmark$ & $\checkmark$ & $\checkmark$  \\
Portfolio Volatility & $\checkmark$ & $\checkmark$ & $\checkmark$ & $\checkmark$ & $\checkmark$  & $\checkmark$  \\ 
\midrule
TP & 1 & 2 & 3 & 4 & 5 & 6\hspace{3mm}  \\ \midrule
SR Net of Turnover &  &  &  &  & $\checkmark$ & $\checkmark$ \\
Sharpe Ratio  &  &  &  &  & $\checkmark$ & $\checkmark$ \\
Portfolio Volatility & $\checkmark$ & $\checkmark$ &  &  & $\checkmark$ &  \\ \bottomrule
\end{tabular}
\end{subtable}\par\bigskip
}
%
\par\bigskip
\scalebox{0.82}{
\begin{tabular}{c}
EWMA \\
\end{tabular}
}\\
\scalebox{0.87}{
\begin{subtable}[t]{0.55\textwidth}
\centering
\begin{tabular}{@{}lllllll@{}}
\toprule
MVP & 1 & 2 & 3 & 4 & 5 & 6 \\ \midrule
SR Net of Turnover &  &  &  &  & $\checkmark$ &  \\
Sharpe Ratio &  &  &  & $\checkmark$* & $\checkmark$ &  \\
Portfolio Volatility & $\checkmark$ & $\checkmark$ & $\checkmark$ & $\checkmark$ & $\checkmark$ & $\checkmark$ \\ 
\midrule
VT  & 1 & 2  & 3 & 4   & 5 & 6 \\ \midrule
SR Net of Turnover & $\times$ & $\checkmark$ &  &  & $\checkmark$ & $\checkmark$ \\
Sharpe Ratio  & $\times$ & $\checkmark$ &  &  & $\checkmark$ & $\checkmark$ \\
Portfolio Volatility & $\checkmark$ & $\checkmark$ & $\checkmark$* &  &  & $\checkmark$ \\ \bottomrule
\end{tabular}%
\end{subtable}
\hfill
\begin{subtable}[t]{0.55\textwidth}
\centering
\begin{tabular}{@{}lllllll@{}}
\toprule
con-MVP & 1 & 2 & 3 & 4 & 5 & 6 \\ \midrule
SR Net of Turnover &  & $\checkmark$ &  &  & $\checkmark$ &  \\
Sharpe Ratio &  & $\checkmark$ &  &  & $\checkmark$ &   \\
Portfolio Volatility & $\checkmark$ & $\checkmark$ & $\checkmark$ & $\checkmark$ & $\checkmark$ & $\checkmark$ \\ 
\midrule
TP & 1 & 2 & 3 & 4 & 5 & 6  \\ \midrule
SR Net of Turnover &  &  &  &  & $\checkmark$ &  \\
Sharpe Ratio  &  &  &  &  & $\checkmark$ &  \\
Portfolio Volatility & $\checkmark$ & $\checkmark$ &  &  & $\checkmark$ & \\ \bottomrule
\end{tabular}
\end{subtable}\par\bigskip
}

\par\bigskip
\scalebox{0.82}{
\begin{tabular}{c}
VAR \\
\end{tabular}
}\\
\scalebox{0.87}{
\begin{subtable}[t]{0.55\textwidth}
\centering
\begin{tabular}{@{}lllllll@{}}
\toprule
MVP & 1 & 2 & 3 & 4 & 5 & 6 \\ \midrule
SR Net of Turnover &  & $\checkmark$ &  & $\checkmark$ & $\checkmark$ & $\checkmark$* \\
Sharpe Ratio & $\times$ & $\checkmark$ &  & $\checkmark$ & $\checkmark$ & $\checkmark$ \\
Portfolio Volatility & $\checkmark$ & $\checkmark$ & $\checkmark$ & $\checkmark$ & $\checkmark$ & $\checkmark$ \\ 
\midrule
VT & 1 & 2  & 3 & 4   & 5 & 6 \\ \midrule
SR Net of Turnover & $\times$ & $\checkmark$ &  & $\checkmark$ & $\checkmark$ & $\checkmark$ \\
Sharpe Ratio  & $\times$ & $\checkmark$ &  & $\checkmark$ & $\checkmark$ & $\checkmark$ \\
Portfolio Volatility & $\checkmark$ & $\checkmark$* &  &  &   &   \\ \bottomrule
\end{tabular}
\end{subtable}
\hfill
\begin{subtable}[t]{0.55\textwidth}
\centering
\begin{tabular}{@{}lllllll@{}}
\toprule
con-MVP & 1 & 2 & 3 & 4 & 5 & 6 \\ \midrule
SR Net of Turnover &  & $\checkmark$* &  & $\checkmark$ & $\checkmark$ & $\checkmark$ \\
Sharpe Ratio &  & $\checkmark$* &  & $\checkmark$ & $\checkmark$ & $\checkmark$ \\
Portfolio Volatility & $\checkmark$ &$\checkmark$  & $\checkmark$ & $\checkmark$ & $\checkmark$ & $\checkmark$ \\ 
\midrule
TP & 1 & 2 & 3 & 4 & 5 & 6  \\ \midrule
SR Net of Turnover &  &  &  &  & $\checkmark$ & $\checkmark$ \\
Sharpe Ratio  &  &  &  &  & $\checkmark$ & $\checkmark$ \\
Portfolio Volatility & $\checkmark$ & $\checkmark$ &  &  & $\checkmark$ &  \\ \bottomrule
\end{tabular}
\end{subtable}\par\bigskip
}

\par\bigskip
\scalebox{0.82}{
\begin{tabular}{c}
VEC \\
\end{tabular}
}\\
\scalebox{0.87}{
\begin{subtable}[t]{0.55\textwidth}
\centering
\begin{tabular}{@{}lllllll@{}}
\toprule
MVP & 1 & 2 & 3 & 4 & 5 & 6 \\ \midrule
SR Net of Turnover & $\times$ & $\checkmark$ &  & $\checkmark$ & $\checkmark$ & $\checkmark$* \\
Sharpe Ratio & $\times$ & $\checkmark$ &  & $\checkmark$ & $\checkmark$ & $\checkmark$ \\
Portfolio Volatility & $\checkmark$ & $\checkmark$ & $\checkmark$ & $\checkmark$ & $\checkmark$ & $\checkmark$ \\ 
\midrule
VT & 1 & 2  & 3 & 4   & 5 & 6 \\ \midrule
SR Net of Turnover & $\times$ & $\checkmark$ &  & $\checkmark$ & $\checkmark$ & $\checkmark$ \\
Sharpe Ratio  & $\times$ & $\checkmark$ &  & $\checkmark$ & $\checkmark$ & $\checkmark$ \\
Portfolio Volatility & $\checkmark$ & $\checkmark$* &  &  &   &   \\ \bottomrule
\end{tabular}
\end{subtable}
\hfill
\begin{subtable}[t]{0.55\textwidth}
\centering
\begin{tabular}{@{}lllllll@{}}
\toprule
con-MVP & 1 & 2 & 3 & 4 & 5 & 6 \\ \midrule
SR Net of Turnover &  & $\checkmark$* &  & $\checkmark$ & $\checkmark$ & $\checkmark$ \\
Sharpe Ratio &  & $\checkmark$* &  & $\checkmark$ & $\checkmark$ & $\checkmark$ \\
Portfolio Volatility & $\checkmark$ &$\checkmark$  & $\checkmark$ & $\checkmark$ & $\checkmark$ & $\checkmark$ \\ 
\midrule
TP & 1 & 2 & 3 & 4 & 5 & 6  \\ \midrule
SR Net of Turnover &  &  &  &  & $\checkmark$ & $\checkmark$ \\
Sharpe Ratio  &  &  &  &  & $\checkmark$ & $\checkmark$ \\
Portfolio Volatility & $\checkmark$ & $\checkmark$ &  & $\times$ & $\checkmark$ &  \\ \bottomrule
\end{tabular}
\end{subtable}\par\bigskip
}
\vspace{0.25cm}
\caption{COV, EWMA, VAR, and VEC}
\scalebox{0.87}{
\begin{tabular}{c}
\multicolumn{1}{p{18cm}}{\textit{Note}: SR Net of Turnover: Sharpe Ratio Net of Turnover Cost. COV, EWMA, VAR, and VEC models are described in Sections~\ref{subsec:cov}--~\ref{subsec:vec} 
and MPV, con-MVP, VT, and TP strategies are described in Section~\ref{sec:pfm}. 
See Table~\ref{tab:mvsr} and Section~\ref{sec:data} for explanations of datasets. Results are for value-weighted data at weekly frequency. $\checkmark$: strategy outperforms naive at 5\% level or lower; $\checkmark$*: strategy outperforms naive at 10\% level; $\times$: strategy underperforms naive; else, insignificant difference.}
\end{tabular}
}
\label{tab:cov}
\end{table}

\begin{table}[htbp!]
\captionsetup{justification=centering,font=small,skip=0pt}
\scalebox{0.82}{
\begin{tabular}{c}
BEKK and ABEKK \\
\end{tabular}
}
\scalebox{0.87}{
\begin{subtable}[t]{0.55\textwidth}
\centering
\begin{tabular}{@{}lllllll@{}}
\toprule
MVP & 1 & 2 & 3 & 4 & 5 & 6 \\ \midrule
SR Net of Turnover &  & $\checkmark$ &  & $\checkmark$ & $\checkmark$ & $\checkmark$ \\
Sharpe Ratio &  & $\checkmark$ &  & $\checkmark$ & $\checkmark$ & $\checkmark$ \\
Portfolio Volatility & $\checkmark$ & $\checkmark$ & $\checkmark$ & $\checkmark$ & $\checkmark$ & $\checkmark$ \\ 
\midrule
VT  & 1 & 2  & 3 & 4   & 5 & 6 \\ \midrule
SR Net of Turnover & $\times$ & $\checkmark$ &  & $\checkmark$ & $\checkmark$ & $\checkmark$\\
Sharpe Ratio  & $\times$ & $\checkmark$ &  & $\checkmark$ & $\checkmark$ & $\checkmark$ \\
Portfolio Volatility & $\checkmark$ & $\checkmark$* &  &  &   &   \\ \bottomrule
\end{tabular}
\end{subtable}
\hfill
\begin{subtable}[t]{0.55\textwidth}
\centering
\begin{tabular}{@{}lllllll@{}}
\toprule
con-MVP & 1 & 2 & 3 & 4 & 5 & 6 \\ \midrule
SR Net of Turnover &  & $\checkmark$* &  & $\checkmark$ & $\checkmark$ & $\checkmark$ \\
Sharpe Ratio &  & $\checkmark$* &  & $\checkmark$ & $\checkmark$ & $\checkmark$ \\
Portfolio Volatility & $\checkmark$ & $\checkmark$ & $\checkmark$ & $\checkmark$ & $\checkmark$ & $\checkmark$ \\ 
\midrule
TP & 1 & 2 & 3 & 4 & 5 & 6  \\ \midrule
SR Net of Turnover &  &  &  &  & $\checkmark$ & $\checkmark$ \\
Sharpe Ratio  &  &  &  &  & $\checkmark$ & $\checkmark$ \\
Portfolio Volatility & $\checkmark$ & $\checkmark$ &  &  & $\checkmark$ &  \\ \bottomrule
\end{tabular}
\end{subtable}\par\bigskip
}

\par\bigskip
\scalebox{0.82}{
\begin{tabular}{c}
CCC \\
\end{tabular}
}
\scalebox{0.87}{
\begin{subtable}[t]{0.55\textwidth}
\centering
\begin{tabular}{@{}lllllll@{}}
\toprule
MVP & 1 & 2 & 3 & 4 & 5 & 6 \\ \midrule
SR Net of Turnover &  & $\checkmark$ &  & $\checkmark$ & $\checkmark$ & $\checkmark$ \\
Sharpe Ratio &  & $\checkmark$ &  & $\checkmark$ & $\checkmark$ & $\checkmark$ \\
Portfolio Volatility & $\checkmark$ & $\checkmark$ & $\checkmark$ & $\checkmark$ & $\checkmark$ & $\checkmark$ \\ 
\midrule
VT & 1 & 2  & 3 & 4   & 5 & 6 \\ \midrule
SR Net of Turnover & $\times$ & $\checkmark$ &  & $\checkmark$ & $\checkmark$ & $\checkmark$\\
Sharpe Ratio  & $\times$ & $\checkmark$ &  & $\checkmark$ & $\checkmark$ & $\checkmark$ \\
Portfolio Volatility & $\checkmark$ & $\checkmark$* &  &  &   &   \\ \bottomrule
\end{tabular}
\end{subtable}
\hfill
\begin{subtable}[t]{0.55\textwidth}
\centering
\begin{tabular}{@{}lllllll@{}}
\toprule
con-MVP & 1 & 2 & 3 & 4 & 5 & 6 \\ \midrule
SR Net of Turnover &  & $\checkmark$ &  & $\checkmark$ & $\checkmark$ & $\checkmark$ \\
Sharpe Ratio &  & $\checkmark$ &  & $\checkmark$ & $\checkmark$ & $\checkmark$ \\
Portfolio Volatility & $\checkmark$ & $\checkmark$ & $\checkmark$ & $\checkmark$ & $\checkmark$ & $\checkmark$ \\ 
\midrule
TP & 1 & 2 & 3 & 4 & 5 & 6  \\ \midrule
SR Net of Turnover &  &  &  &  & $\checkmark$ & $\checkmark$ \\
Sharpe Ratio  &  &  &  &  & $\checkmark$ & $\checkmark$ \\
Portfolio Volatility & $\checkmark$ & $\checkmark$ &  &  & $\checkmark$ &  \\ \bottomrule
\end{tabular}
\end{subtable}\par\bigskip
}

\par\bigskip
\scalebox{0.82}{
\begin{tabular}{c}
DCC and ADCC \\
\end{tabular}
}
\scalebox{0.87}{
\begin{subtable}[t]{0.55\textwidth}
\centering
\begin{tabular}{@{}lllllll@{}}
\toprule
MVP & 1 & 2 & 3 & 4 & 5 & 6 \\ \midrule
SR Net of Turnover &  & $\checkmark$ &  & $\checkmark$ & $\checkmark$ & $\checkmark$ \\
Sharpe Ratio &  & $\checkmark$ &  & $\checkmark$ & $\checkmark$ & $\checkmark$ \\
Portfolio Volatility & $\checkmark$ & $\checkmark$ & $\checkmark$ & $\checkmark$ & $\checkmark$ & $\checkmark$ \\ 
\midrule
VT  & 1 & 2  & 3 & 4   & 5 & 6 \\ \midrule
SR Net of Turnover & $\times$ & $\checkmark$ &  & $\checkmark$ & $\checkmark$ & $\checkmark$\\
Sharpe Ratio  & $\times$ & $\checkmark$ &  & $\checkmark$ & $\checkmark$ & $\checkmark$ \\
Portfolio Volatility & $\checkmark$ & $\checkmark$* &  &  &   &   \\ \bottomrule
\end{tabular}
\end{subtable}
\hfill
\begin{subtable}[t]{0.55\textwidth}
\centering
\begin{tabular}{@{}lllllll@{}}
\toprule
con-MVP & 1 & 2 & 3 & 4 & 5 & 6 \\ \midrule
SR Net of Turnover &  & $\checkmark$* &  & $\checkmark$ & $\checkmark$ & $\checkmark$ \\
Sharpe Ratio &  & $\checkmark$* &  & $\checkmark$ & $\checkmark$ & $\checkmark$ \\
Portfolio Volatility & $\checkmark$ & $\checkmark$ & $\checkmark$ & $\checkmark$ & $\checkmark$ & $\checkmark$ \\ 
\midrule
TP & 1 & 2 & 3 & 4 & 5 & 6  \\ \midrule
SR Net of Turnover &  &  &  &  & $\checkmark$ & $\checkmark$ \\
Sharpe Ratio  &  &  &  &  & $\checkmark$ & $\checkmark$ \\
Portfolio Volatility & $\checkmark$ & $\checkmark$ &  &  & $\checkmark$ &  \\ \bottomrule
\end{tabular}
\end{subtable}\par\bigskip
}
\par\bigskip
\scalebox{0.82}{
\begin{tabular}{c}
COPULA \\
\end{tabular}
}
\scalebox{0.87}{
\begin{subtable}[t]{0.55\textwidth}
\centering
\begin{tabular}{@{}lllllll@{}}
\toprule
MVP & 1 & 2 & 3 & 4 & 5 & 6 \\ \midrule
SR Net of Turnover &  & $\checkmark$ &  & $\checkmark$* & $\checkmark$ &  \\
Sharpe Ratio &  & $\checkmark$ &  & $\checkmark$* & $\checkmark$ &  \\
Portfolio Volatility & $\checkmark$ & $\checkmark$ & $\checkmark$ & $\checkmark$ & $\checkmark$ & $\checkmark$ \\ 
\midrule
VT  & 1 & 2  & 3 & 4   & 5 & 6 \\ \midrule
SR Net of Turnover & $\times$ & $\checkmark$ &  & $\checkmark$ & $\checkmark$ & $\checkmark$ \\
Sharpe Ratio  & $\times$ & $\checkmark$ &  & $\checkmark$ & $\checkmark$ & $\checkmark$ \\
Portfolio Volatility & $\checkmark$ & $\checkmark$* &  &  &  & \\ \bottomrule
\end{tabular}%
\end{subtable}
\hfill
\begin{subtable}[t]{0.55\textwidth}
\centering
\begin{tabular}{@{}lllllll@{}}
\toprule
con-MVP & 1 & 2 & 3 & 4 & 5 & 6 \\ \midrule
SR Net of Turnover &  & $\checkmark$ &  & $\checkmark$ & $\checkmark$ &  \\
Sharpe Ratio &  & $\checkmark$ &  & $\checkmark$ & $\checkmark$ &   \\
Portfolio Volatility & $\checkmark$ & $\checkmark$ & $\checkmark$ & $\checkmark$ & $\checkmark$ & $\checkmark$*\\ 
\midrule
TP & 1 & 2 & 3 & 4 & 5 & 6  \\ \midrule
SR Net of Turnover &  &  &  &  & $\checkmark$ &  \\
Sharpe Ratio  &  &  &  &  & $\checkmark$ &  \\
Portfolio Volatility & $\checkmark$ & $\checkmark$ &  & $\checkmark$ & $\checkmark$ & $\checkmark$ \\ \bottomrule
\end{tabular}
\end{subtable}\par\bigskip
}

\vspace{0.25cm}
\caption{BEKK, ABEKK, CCC, DCC, ADCC, and COPULA}
\scalebox{0.87}{
\begin{tabular}{c}
\multicolumn{1}{p{18cm}}{\textit{Note}: SR Net of Turnover: Sharpe Ratio Net of Turnover Cost. BEKK, ABEKK, CCC, DCC, ADCC, and COPULA models are described in Sections~\ref{subsec:bekk}--\ref{subsec:copula} 
and MPV, con-MVP, VT, and TP strategies are described in Section~\ref{sec:pfm}. See Table~\ref{tab:mvsr} and Section~\ref{sec:data} for explanations of datasets.  
Results are for value-weighted data at weekly frequency. $\checkmark$: strategy outperforms naive at 5\% level or lower; $\checkmark$*: strategy outperforms naive at 10\% level; $\times$: strategy underperforms naive; else, insignificant difference.}
\end{tabular}
}
\label{tab:vec}
\end{table}

\begin{table}[htbp!]
\captionsetup{justification=centering,font=small,skip=0pt}
\scalebox{0.82}{
\begin{tabular}{c}
RSVAR \\
\end{tabular}
}
\scalebox{0.87}{
\begin{subtable}[t]{0.55\textwidth}
\centering
\begin{tabular}{@{}lllllll@{}}
\toprule
MVP & 1 & 2 & 3 & 4 & 5 & 6 \\ \midrule
SR Net of Turnover & $\times$ &  &  & $\checkmark$ & $\checkmark$ & $\checkmark$ \\
Sharpe Ratio & $\times$ &  &  & $\checkmark$ & $\checkmark$ & $\checkmark$ \\
Portfolio Volatility & $\checkmark$ &  & & $\checkmark$* &  &  \\ 
\midrule
VT  & 1 & 2  & 3 & 4   & 5 & 6 \\ \midrule
SR Net of Turnover & $\times$ &  &  & $\checkmark$ & $\checkmark$ & $\checkmark$ \\
Sharpe Ratio  & $\times$ &  &  & $\checkmark$ & $\checkmark$ & $\checkmark$ \\
Portfolio Volatility & $\checkmark$ &  &  & $\checkmark$* &   & $\checkmark$* \\ \bottomrule
\end{tabular}%
\end{subtable}
\hfill
\begin{subtable}[t]{0.55\textwidth}
\centering
\begin{tabular}{@{}lllllll@{}}
\toprule
con-MVP & 1 & 2 & 3 & 4 & 5 & 6 \\ \midrule
SR Net of Turnover &  &  &  & $\checkmark$ & $\checkmark$ & $\checkmark$ \\
Sharpe Ratio & $\times$ &  &  & $\checkmark$ & $\checkmark$ & $\checkmark$  \\
Portfolio Volatility & $\checkmark$ &  &  & $\checkmark$* &   &   \\ 
\midrule
TP & 1 & 2 & 3 & 4 & 5 & 6  \\ \midrule
SR Net of Turnover &  &  &  & $\checkmark$ & $\checkmark$ & $\checkmark$ \\
Sharpe Ratio  &  &  &  & $\checkmark$ & $\checkmark$ & $\checkmark$ \\
Portfolio Volatility & $\checkmark$ &  &  & $\checkmark$* &  &  \\ \bottomrule
\end{tabular}
\end{subtable}\par\bigskip
}
\par\bigskip
\scalebox{0.82}{
\begin{tabular}{c}
MSV \\
\end{tabular}
}
\scalebox{0.87}{
\begin{subtable}[t]{0.55\textwidth}
\centering
\begin{tabular}{@{}lllllll@{}}
\toprule
MVP & 1 & 2 & 3 & 4 & 5 & 6 \\ \midrule
SR Net of Turnover &  &  &  & $\checkmark$* & $\checkmark$ &  \\
Sharpe Ratio &  &  &  & $\checkmark$* & $\checkmark$ &  \\
Portfolio Volatility & $\checkmark$ & $\checkmark$ & $\checkmark$ & $\checkmark$ & $\checkmark$ & $\checkmark$ \\ 
\midrule
VT  & 1 & 2  & 3 & 4   & 5 & 6 \\ \midrule
SR Net of Turnover & $\times$ & $\checkmark$ & $\checkmark$* & $\checkmark$* & $\checkmark$ & $\checkmark$ \\
Sharpe Ratio  & $\times$ & $\checkmark$ & $\checkmark$* & $\checkmark$* & $\checkmark$ & $\checkmark$ \\
Portfolio Volatility & $\checkmark$ & $\checkmark$ &  &  &   & $\checkmark$ \\ \bottomrule
\end{tabular}%
\end{subtable}
\hfill
\begin{subtable}[t]{0.55\textwidth}
\centering
\begin{tabular}{@{}lllllll@{}}
\toprule
con-MVP & 1 & 2 & 3 & 4 & 5 & 6 \\ \midrule
SR Net of Turnover &  & $\checkmark$* &  & $\checkmark$ & $\checkmark$ & $\checkmark$* \\
Sharpe Ratio &  & $\checkmark$* &  & $\checkmark$ & $\checkmark$ & $\checkmark$*  \\
Portfolio Volatility & $\checkmark$ & $\checkmark$ & $\checkmark$ & $\checkmark$ & $\checkmark$ & $\checkmark$  \\ 
\midrule
TP & 1 & 2 & 3 & 4 & 5 & 6  \\ \midrule
SR Net of Turnover &  &  &  &  & $\checkmark$ & $\checkmark$* \\
Sharpe Ratio  &  &  &  & $\checkmark$* & $\checkmark$ & $\checkmark$* \\
Portfolio Volatility & $\checkmark$ & $\checkmark$ &  &  & $\checkmark$ & $\checkmark$ \\ \bottomrule
\end{tabular}
\end{subtable}\par\bigskip
}
\par\bigskip
\scalebox{0.82}{
\begin{tabular}{c}
RCOV \\
\end{tabular}
}
\scalebox{0.87}{
\begin{subtable}[t]{0.55\textwidth}
\centering
\begin{tabular}{@{}lllllll@{}}
\toprule
MVP & 1 & 2 & 3 & 4 & 5 & 6 \\ \midrule
SR Net of Turnover &  & $\checkmark$* &  & $\checkmark$ & $\checkmark$ &  \\
Sharpe Ratio &  & $\checkmark$* &  & $\checkmark$ & $\checkmark$ &  \\
Portfolio Volatility & $\checkmark$ & $\checkmark$ & $\checkmark$ & $\checkmark$ & $\checkmark$ & $\checkmark$ \\ 
\midrule
VT  & 1 & 2  & 3 & 4   & 5 & 6 \\ \midrule
SR Net of Turnover &  & $\checkmark$ & $\checkmark$* & $\checkmark$ & $\checkmark$ & $\checkmark$ \\
Sharpe Ratio  &  & $\checkmark$ & $\checkmark$* & $\checkmark$ & $\checkmark$ & $\checkmark$ \\
Portfolio Volatility & $\checkmark$ & $\checkmark$ &  &  &   &  \\ \bottomrule
\end{tabular}
\end{subtable}
\hfill
\begin{subtable}[t]{0.55\textwidth}
\centering
\begin{tabular}{@{}lllllll@{}}
\toprule
con-MVP & 1 & 2 & 3 & 4 & 5 & 6 \\ \midrule
SR Net of Turnover &  &  &  & $\checkmark$ & $\checkmark$ & $\checkmark$ \\
Sharpe Ratio &  &  &  & $\checkmark$ & $\checkmark$ & $\checkmark$ \\
Portfolio Volatility & $\checkmark$ & $\checkmark$ & $\checkmark$ & $\checkmark$ & $\checkmark$  & $\checkmark$  \\ 
\midrule
TP & 1 & 2 & 3 & 4 & 5 & 6  \\ \midrule
SR Net of Turnover &  & $\checkmark$ & $\checkmark$ &  & $\checkmark$ & $\checkmark$ \\
Sharpe Ratio  &  & $\checkmark$ & $\checkmark$ &  & $\checkmark$ & $\checkmark$ \\
Portfolio Volatility & $\checkmark$ & $\checkmark$ &  & $\times$ & $\checkmark$ & $\checkmark$ \\ \bottomrule
\end{tabular}
\end{subtable}\par\bigskip
}
\par\bigskip
\scalebox{0.82}{
\begin{tabular}{c}
CP \\
\end{tabular}
}
\scalebox{0.87}{
\begin{subtable}[t]{0.55\textwidth}
\centering
\begin{tabular}{@{}lllllll@{}}
\toprule
MVP & 1 & 2 & 3 & 4 & 5 & 6 \\ \midrule
SR Net of Turnover &  & $\checkmark$* &  & $\checkmark$ & $\checkmark$ & $\checkmark$ \\
Sharpe Ratio &  & $\checkmark$ &  & $\checkmark$ & $\checkmark$ & $\checkmark$ \\
Portfolio Volatility & $\checkmark$ &  & & $\checkmark$ & $\checkmark$* & $\checkmark$ \\ 
\midrule
VT  & 1 & 2  & 3 & 4   & 5 & 6 \\ \midrule
SR Net of Turnover & $\times$ &  &  & $\checkmark$ & $\checkmark$ & $\checkmark$ \\
Sharpe Ratio  & $\times$ &  &  & $\checkmark$ & $\checkmark$ & $\checkmark$ \\
Portfolio Volatility & $\checkmark$ &  &  & $\checkmark$ &   &   \\ \bottomrule
\end{tabular}%
\end{subtable}
\hfill
\begin{subtable}[t]{0.55\textwidth}
\centering
\begin{tabular}{@{}lllllll@{}}
\toprule
con-MVP & 1 & 2 & 3 & 4 & 5 & 6 \\ \midrule
SR Net of Turnover &  & $\checkmark$ &  & $\checkmark$ & $\checkmark$ & $\checkmark$ \\
Sharpe Ratio &  & $\checkmark$ &  & $\checkmark$ & $\checkmark$ & $\checkmark$  \\
Portfolio Volatility & $\checkmark$ &  &  & $\checkmark$ & $\checkmark$*  & $\checkmark$*  \\ 
\midrule
TP & 1 & 2 & 3 & 4 & 5 & 6  \\ \midrule
SR Net of Turnover &  &  &  & $\checkmark$ & $\checkmark$ & $\checkmark$ \\
Sharpe Ratio  &  &  &  & $\checkmark$ &  $\checkmark$ & $\checkmark$ \\
Portfolio Volatility & $\checkmark$ &  &  & $\checkmark$ &  & $\checkmark$ \\ \bottomrule
\end{tabular}
\end{subtable}\par\bigskip
}
\vspace{0.25cm}
\caption{RSVAR, MSV, RCOV, and CP}
\scalebox{0.87}{
\begin{tabular}{c}
\multicolumn{1}{p{18cm}}{\textit{Note}: SR Net of Turnover: Sharpe Ratio Net of Turnover Cost. MSV, RCOV, and CP models are described in Sections~\ref{subsec:rsvar}--\ref{subsec:cp} 
and MPV, con-MVP, VT, and TP strategies are described in Section~\ref{sec:pfm}. Section~\ref{sec:metrics} for explanations of econometric model abbreviations. See Table~\ref{tab:mvsr} and Section~\ref{sec:data} for explanations of datasets. Results are for value-weighted data at weekly frequency. $\checkmark$: strategy outperforms naive at 5\% level or lower; $\checkmark$*: strategy outperforms naive at 10\% level; $\times$: strategy underperforms naive; else, insignificant difference.}
\end{tabular}
}
\label{tab:msv}
\end{table}

\end{document}